\documentclass[sigconf,authorversion,nonacm,balance=false]{acmart}
\usepackage{url}
\PassOptionsToPackage{hyphens}{url}\usepackage{hyperref}

\usepackage[utf8]{inputenc}

\usepackage{tabularray}
\usepackage{tabularx}

\usepackage{enumitem} 

\usepackage{graphicx}

\usepackage{csquotes}
\newcommand{\icite}[1]{[#1]}
\let\oldtextquote\textquote
\renewcommand{\textquote}[1]{{\itshape\oldtextquote{#1}}}
\renewcommand{\mkbegdispquote}[2]{\itshape}

\usepackage{datetime}

\yyyymmdddate

\begin{document}
	
\setcitestyle{citesep={;}, notesep={, }}

\title{The PET Paradox}
\subtitle{How Amazon Instrumentalises PETs in Sidewalk to Entrench Its Infrastructural Power}

\author{Thijmen van Gend (t.n.vangend@tudelft.nl)}
\orcid{0000-0001-9157-9107}

\author{Donald Jay Bertulfo (d.j.bertulfo@tudelft.nl)}
\orcid{0000-0002-5459-7661}

\author{Seda Gürses (f.s.gurses@tudelft.nl)}
\orcid{0000-0003-3178-8710}
\affiliation{
  \institution{Delft University of Technology (faculty of Technology, Policy and Management)}
\country{The Netherlands}  \\
 	  \textit{\textbf{This is a preprint last edited on \today\ and currently under review. We welcome all comments and feedback.}}
 }

\renewcommand{\shortauthors}{Van Gend et al.}

\begin{abstract} 
Recent applications of Privacy Enhancing Technologies (PETs) reveal a paradox. PETs aim to alleviate power asymmetries, but can actually entrench the infrastructural power of companies implementing them vis-à-vis other public and private organisations. We investigate whether and how this contradiction manifests with an empirical study of Amazon's cloud connectivity service called Sidewalk.
In 2021, Amazon remotely updated Echo and Ring devices in consumers' homes, to transform them into Sidewalk \textquote{gateways}. Compatible Internet of Things (IoT) devices, called \textquote{endpoints}, can connect to an associated \textquote{Application Server} in Amazon Web Services (AWS) through these gateways. We find that Sidewalk is not just a connectivity service, but an extension of Amazon's cloud infrastructure as a software production environment for IoT manufacturers. PETs play a prominent role in this pursuit: we observe a two-faceted PET paradox. First, suppressing some information flows allows Amazon to promise narrow privacy guarantees to owners of Echo and Ring devices when \textquote{flipping} them  into gateways. Once flipped, these gateways constitute a crowdsourced connectivity infrastructure that covers 90\% of the US population and expands their AWS offerings. We show how novel information flows, enabled by Sidewalk connectivity, raise greater surveillance and competition concerns. Second, Amazon governs the implementation of these PETs, requiring manufacturers to adjust their device hardware, operating system and software; cloud use; factory lines; and organisational processes. Together, these changes turn manufacturers' endpoints into accessories of Amazon's computational infrastructure; further entrenching Amazon's infrastructural power. We argue that power analyses undergirding PET design should go beyond analysing information flows. We propose future steps for policy and tech research. 
\end{abstract}


\maketitle


\section{Introduction}\label{sec_introduction}
PETs are a family of technologies that aim to provide mathematically undergirded privacy guarantees to people vis-à-vis organisations that collect and process data. They do so by putting strict controls on data collection and processing that may otherwise lead to infringements of individual privacy and power asymmetries. Both aspirations are evident in how PET designers strictly interpret data minimisation to the purpose of processing, and the community's commitment to developing distributed designs that avoid single points of (privacy) failure \cite{gursesEngineeringPrivacyDesign2015}. More broadly, actors and their potential to overreach using various sources of power (e.g. legal) prevalently feature in adversarial and threat models, underlining the importance of power asymmetries in designing PETs.

The output of a vibrant community of researchers and practitioners convening since the 1990s, PETs adoption remained marginal until the beginning of the 2010s~\cite{goldbergPrivacyEnhancingTechnologiesInternet2007}. This contrasted with the popularity of other privacy technologies that entrust service providers with the protection of data~\cite{gursesPrivacyEngineeringShaping2016, gursesTwoTalesPrivacy2013, spiekermannEngineeringPrivacy2008}. Two pivotal moments helped change the status quo. The Snowden revelations increased awareness of large-scale surveillance programs, and incentivised experts to develop and deploy PETs more widely ~\cite{ungerSoKSecureMessaging2015, farrellReflectionsTenYears2023, barnesConfidentialityFacePervasive2015}. Companies also responded: Meta/WhatsApp \cite{greenbergWhatsappJustSwitched2014} and Apple \cite{apuzzoAppleOtherTech2015} adopted PETs, bringing PETs to large swaths of users. Second, in adopting the General Data Protection Regulation, the EU codified privacy by design as an approach to curbing informational power asymmetries \cite[e.g.][]{informationcommissionersofficePrivacyenhancingTechnologiesPETs2023, europeanunionagencyforcybersecurityDataProtectionEngineering2022, sroujiHowPrivacyenhancingTechnologies2020, ausloosFoundationsDataProtection2020}. Critics have since argued that PETs also allowed companies to strategically manage their response to transnational surveillance~\cite{vanhobokenPrivacySecurityCloud2014} and reduce their data protection obligations~\cite{vealeDeniedDesignData2023}. Yet, however strategic PET adoption may be, these developments generally confirmed that PETs can deliver control over information flows to address power asymmetries vis-à-vis companies and governments, as well as between them.
	
A decade in, strategic use of PETs increasingly challenges their promise to push back on power \textquote{by design}. Especially concerning is when companies with broad control over the functionality of clients -- e.g. through operating systems (OSes) and web browsers -- deploy PETs to constrain information flows in a way that provides privacy to clients (in a business-to-consumer (B2C) context), while empowering themselves vis-à-vis consumers, and entrenching their role in the production and delivery of digital services (business-to-government (B2G) and -business (B2B)). We see such examples by Google and Apple in contact tracing, device finding, and advertising (see §\ref{sec_PETs_as_a_source_of_power}). These point to shifts in the power game.

In this paper, we present an empirical study of one such case: Amazon Sidewalk. We investigate how Amazon instrumentalises PETs in entrenching its cloud environment vis-à-vis Internet of Things (IoT) manufacturers. Sidewalk is a crowdsourced networking service where \textquote{gateways} (Amazon Echo and Ring consumer devices) share a portion of their wifi bandwidth with other IoT devices (\textquote{endpoints}, potentially owned by others). The latter can then connect to their manufacturer's \textquote{Application Server} through and in Amazon's cloud (Amazon Web Services (AWS)) \cite{amazonAmazonSidewalkn.d.}.

Sidewalk builds on Amazon's existing control over consumer devices, transforming them into gateways, and therewith extensions of AWS. Echo and Ring devices are typically considered devices for personal use. Thanks to how large-scale software is produced today, these \textquote{personal} devices are better considered \textquote{accessories of Amazon's cloud}: the devices are key to software development and deployment using AWS, and are increasingly managed through the clouds. 
Hence, unlike personal computers that gave consumers some autonomy and freedom to tinker~\cite{keltyFogFreedom2014}, users now purchase and maintain IoT devices that Amazon uses to deliver and improve services \textit{to them}. These services include Amazon's own as well as third party services (\textquote{Skills}) on Alexa. As a result, users rely on a connection to AWS for their devices to function; their \textquote{freedom to tinker} reduced to controlling certain device settings. Amazon, however, goes further and transforms these devices from accessories to extensions of their cloud, with the help of PETs. With PETs (i.e. encryption and obfuscation), Sidewalk promises some privacy guarantees to users in order to \textquote{flip} Echo and Ring devices into Sidewalk gateways. Endpoints can then connect to the cloud \textit{through} these gateways. This transforms the Echo and Ring devices into an actual extension of Amazon's cloud; constituting what we call \textit{computational infrastructures} (clouds plus end devices) for software production that can be used by AWS clients.

Our study traces the role of PETs in the extension of AWS as software \emph{production environment} to IoT manufacturers, and argues Sidewalk is not just a connectivity service.
We show that the freshly transformed gateways connect IoT devices manufactured by others to Sidewalk and thus to Amazon's cloud; rendering them accessories of Amazon's cloud, too.
Sidewalk provides a window into the myriad of ways in which Amazon leverages its infrastructural position in the implementation of PETs, reshaping B2B relationships and centring its cloud services in businesses' production of services in the process. 
As we will show, Sidewalk's use of PETs has implications not only for endpoint and gateway users (i.e. their privacy; B2C), but also, if not more, for endpoint manufacturers (concerning how they produce these endpoints; B2B).

We use qualitative empirical methods (§\ref{sec_methods_and_case_description}) to answer our main research question: \textit{Whether and how does Amazon instrumentalise PETs in entrenching its computational infrastructure as a production environment vis-à-vis IoT manufacturers?} To contextualise Amazon's efforts and increase our analytical precision when using terms such as privacy and power, we first survey other instances wherein companies with infrastructural control operationalised PETs, and the power it yielded them (§\ref{sec_PETs_as_a_source_of_power}). These instances demonstrate concrete ways in which the particular implementation of data minimisation, purpose limitation, and anonymisation through PETs can backfire on its promise to mitigate power asymmetries in different constellations of infrastructural control. Subsequently, we ask: \textit{How does Amazon's implementation of PETs shape user privacy and Sidewalk's control over information flows?} (§\ref{sec_information_flows_and_PETs_in_sidewalk}). This part starts from a consumer-oriented (B2C) approach to the use of PETs, and expands to address greater information flows in Sidewalk, including their impact on manufacturers (B2B). We skip a full-blown privacy analysis in favour of making space for a broader power analysis. Next, we ask: \textit{How does Amazon's implementation and governance of PETs in Sidewalk affect manufacturers' production of IoT devices and services?} (§\ref{sec_PETs_transforming_production}). Herein, we adopt a B2B-oriented \textquote{production view}, i.e., we pay specific attention to how adopting Sidewalk affects how IoT manufacturers produce their devices and services; and accordingly, how power dynamics emerge between them and Amazon. To do so, we centre the actors; physical and digital inputs, outputs, and processes; and (software and hardware) production environments involved in producing Sidewalk and IoT services. In the discussion (§\ref{sec_discussion}), we argue how these power dynamics lead to the \textquote{PET Paradox}. After exploring limitations (§\ref{sec_limitations}), we conclude with the questions that our analysis raises for academia and policy, and propose paths forward (§\ref{sec_conclusion}). We cannot answer all these questions, but contribute a rich account of how PETs are instrumentalised in engineered environments as a ground to start discussing them.

\section{PETs as a source of power}\label{sec_PETs_as_a_source_of_power}
Over the last decade, more and more tech companies have included PETs in their repertoires. Their rise to prominence is also evident in the PET Symposium programme of the last decade. While we can consider these positive developments, tech companies may leverage PETs strategically, e.g. to claim that certain regulations no longer apply when they use PETs \cite{woodsLitigatingDataSovereignty2018, vealeDeniedDesignData2023}. Some of this is due to the underlying protection mechanisms: moderating content or providing data access to data subjects, researchers, and public bodies may be hard when said data is encrypted, which led to contentious use of PETs for client side scanning~\cite{abelsonBugsOurPockets2024}.

We focus on recent large-scale implementations that have raised further concerns with respect to PETs being used not only to protect privacy, but also to entrench the market and infrastructural position of these already powerful players in the production of services by other actors. We discuss three instances that demonstrate how, in doing so, they increase their influence in public service delivery (i.e. digital contact tracing), bolster their market position (i.e. device finding networks), and remain central to digital advertising. \\

\noindent \textbf{Instance 1: Digital Contact Tracing.} During the Covid-19 pandemic, Google and Apple leveraged PETs to position themselves as arbiters of citizens' privacy vis-à-vis health authorities and researchers for digital contact tracing (DCT). Accordingly, the companies shaped how they could \textquote{produce} public health services.

As the pandemic unfolded at great speed and scale, governments called for taking advantage of high penetration numbers of \textquote{smart phones} for a key public health service: contact tracing. Various coalitions from industry and academia responded to the call. One such coalition proposed the Decentralised Privacy-Preserving Proximity Tracking (DP3T) protocol, that aimed to maximise privacy preservation and purpose limitation \cite{troncosoDeployingDecentralizedPrivacypreserving2022a}. DP3T relied on persistent background functioning, that operating systems (OS) prohibit for privacy and performance reasons. This required Google and Apple, that control 99\% of the smartphone OSes in the concerned regions \cite{statcounterMobileOperatingSystem2024}, to adjust their OSes.

Amidst a greater controversy around decentralised solutions prioritising privacy over information flowing to public health authorities \cite{JointStatementContact2020}, the companies came forth with an implementation of DP3T in their OSes. This they called the \textquote{Google and Apple Exposure Notification} (GAEN) framework. GAEN exposed APIs that contact tracing apps developed by health authorities could invoke. However, these APIs presented health authorities with a \textquote{heavily constrained set of parameters}, which \textquote{strongly limited the design choices of app developers in making tradeoffs among privacy, security, and epidemiological utility of the applications} (p. 53). To illustrate: in the name of user privacy, the API would initially only expose highly summarised information, hampering calculation of daily viral exposure accumulation by epidemiologists. By pushing the protocol into their OSes, Google and Apple positioned themselves as public health infrastructure providers; resolved a public controversy around PETs on their own terms; and secured seats in the public health decision-making arena. \\

\noindent \textbf{Instance 2: Device finding networks.} With an opt-out update \cite{fleishmanHowOptOut2021}, Apple turned \textquote{hundreds of millions} \cite{appleFindMyNetworkn.d.} Apple products into a crowdsourced privacy-preserving infrastructure for finding devices, called Find My. This feat cannot be pulled off by other tracking companies that do not control a smartphone OS, yet was found to compromise their competitive position.

Find My-enabled devices report the location of nearby devices to their owners \cite{appleAppleFindMy2021, appleAppleIntroducesAirTag2021}. Compatible devices include Apple's own devices and asset trackers (AirTags), but also third-party trackers and products (e.g. earphones) \cite{boweAllNonAppleGadgets2022, menonBestFindMy2023}). Third-party tracker manufacturers typically have smaller networks and rely on customers downloading a smartphone app, to both manage their own tracker and report the location of others' \cite{tileHowTileWorksn.d., chipoloChipolon.d., pebblebeePebblebeen.d.}. For instance, \citeauthor{chipoloChipolon.d.}'s network counts 5 million contributing devices \cite{chipoloChipolon.d.}.

Apple allows other businesses to make Find My-compatible products, under potentially unfair constraints. Adopting companies must join Apple's developer programs and pass a certification process \cite{appleFindMyNetworkn.d.}. Their devices may not support Apple's and their own network simultaneously \cite{daruTestimonyKirstenDaru2021, pebblebeeClipCardTagn.d.}. According to an unofficial publication of the Find My specification, this \textquote{may interfere with the security and privacy requirements} \cite[][p. 14]{appleFindMyNetwork2020}. Granted Apple's ability to remotely turn consumer-owned Apple devices into \textquote{finders}, adopting Find My and subjecting themselves to its policies may feel inevitable for other tracker manufacturers ~\cite{brownleeAppleVsParadox2021}.
These conditions for adoption were investigated by antitrust authorities worldwide \cite{espinozaAppleAccusedCompetition2020, tileObservationsTileStatement2021, competitionandmarketsauthorityMobileEcosystemsMarket2022, unitedstatessenatecommitteeonthejudiciaryAntitrustAppliedExamining2021, potuckAppleAntitrustHearing2021, robins-earlyUSAccusesApple2024, daruTestimonyKirstenDaru2021}, alleging Apple of gatekeeping and self-preferencing while weaponising security and privacy. \\

\noindent \textbf{Instance 3: Digital advertising.} In this case, efforts by Apple and Google aimed to make digital advertising more privacy-enhancing, entrenched their position in the business models and production processes of companies in the digital advertising space.

In 2021, Apple implemented the \textquote{App Tracking Transparency} (ATT) framework in their mobile OS. Consequently, app developers must ask user consent to use device identifiers for cross-app tracking \cite{appleIfAppAsks2024}. Estimates of opt-in rates vary widely between 11\% and 50\% \cite{curryAppTrackingTransparency2024, mackenzieHeyBigSpender2024, wetzlerATTTwoYears2023, appsflyerAppsFlyerDataReveals2024}. As privacy-enhancing alternative for ad attribution, Apple offers the SKAdNetwork framework. SKAdNetwork uses on-device processing and delays and anonymises attribution reporting with advertisers \cite{mcguiganPrivateAttributesMeanings2023}.

ATT seriously hurt the revenue of Meta \cite{congerChangeAppleTormenting2022} and had a small effect of steering app developers from advertising revenue models to paid apps and in-app payments \cite{keslerImpactAppleApp2023}. Meanwhile, Apple's advertising market share tripled \cite{mcgeeApplePrivacyChanges2021}. The UK competition Authority [\citealp[][pp. 233-244]{competitionandmarketsauthorityMobileEcosystemsMarket2022}] and \citet{mcguiganPartyCynicalResignation2023} claim this could be because (1) Apple exempts their own tracking of users across Apple apps from ATT, claiming it constitutes first-party data; (2) their \textquote{Personalised Ads} nudge users to opt in, whereas ATT pop-ups nudge users to opt out; and (3) Apple uses the Ads Attribution API for Apple ads, while other app developers and ad networks must make do with the SKAdNetwork APIs, that yields advertisers delayed and less granular information. Accordingly, Apple can track user behaviour more accurately and provide advertisers with richer ad performance insights than competing ad networks. Competition and consumer authorities in multiple countries have initiated (presently ongoing) investigations into whether ATT increases barriers to entry and enables self-preferencing	\cite{autoritedelaconcurrenceTargetedAdvertisingApple2021, autoritedelaconcurrenceAdvertisingIOSMobile2023, bundeskartellamtBundeskartellamtReviewsApple2022, autoritagarantedellaconcorrenzaedelmercatoA561A561BItalianCompetition2023, urzadochronykonkurencjiikonsumentowApplePresidentUOKiK2021, competitionandmarketsauthorityMobileEcosystemsMarket2022}. \\
	
\noindent In early 2020, Google announced \textquote{Privacy Sandbox} (PS) for Chrome and Android. For Chrome, PS would deprecate third-party cookies, but still allow first-party cookies \cite{schuhBuildingMorePrivate2020}. PS includes APIs to substitute third-party cookies' functionalities in an allegedly more privacy-preserving way, with decentralised PETs that run inside users' browsers \cite{googleWhatPrivacySandbox2024}. In fact, Google's instrumentalisation of PETs entrenches their role in online advertising, aggravating dependencies of advertisers, publishers, and adtech providers on them.

While deprecating third-party cookies could be a win for user privacy \cite{bonnerICOStatementResponse2024}, 
governmental authorities have started investigations
[\citealp{commissionnationaledelinformatiqueetdeslibertesPubliciteLigneCNIL2024}; \citealp[][pp. G123-124]{competitionandmarketsauthorityAppendixRoleTracking2020}; \citealp[][p. 296]{competitionandmarketsauthorityOnlinePlatformsDigital2020}; \citealp{competitionandmarketsauthorityCase50972Privacy2022}; \citealp[][pp. 17-18, 31-32]{competitionandmarketsauthorityCMAQ120242024}; \citealp{competitionandmarketsauthorityInvestigationGooglePrivacy2024}; \citealp{poledexpertisedelaregulationnumeriquePrivacySandboxCollection2022}] arguing that Google disproportionally disadvantages publishers and adtech companies.
The PS APIs are less effective than third-party cookies \cite{commissionnationaledelinformatiqueetdeslibertesPubliciteLigneCNIL2024}, while costing publishers ad revenue, and necessitating a technologically complex and expensive implementation \cite{parsonsPrivacySandboxTesting2024, indexexchangeInsightsOurPrivacy2024}. Alternatives to PS for publishers suffer similar complexity, effectiveness, and cost issues. Further, Google may benefit from publishers flocking to its advertising services (e.g. Google Ad Manager \cite{parsonsPrivacySandboxTesting2024}); and from advertisers moving to closed ecosystems with better tracking capabilities (e.g. mobile, search, and social media) \cite{titoneWhyGoogleNo2024, titoneAssessingImpactGoogle2024}, where Google allegedly has higher margins \cite{seufertWhyGoogleKilling2024, dotanHereRankingGoogle2022}. Google's dual role as adtech company and as the party defining how adtech can work in Chrome, also raises self-preferencing concerns.\footnote{Google announced in July 2024 a pivot to a user consent model, mentioning the UK competition authority's investigation \cite{chavezNewPathPrivacy2024}. Google not communicating a timeline nor details about the opt-out design upset publishers, creating uncertainty on what level of third-party cookie adoption to expect and by when \cite{barberWhyGoogleCookie2024, josephGoogleAssuresAd2024}; especially affecting smaller publishers with less resources to invest in testing alternatives \cite{shieldsAdTechBosses2024, josephGoogleAssuresAd2024}.} \\

\noindent Arguably, it is not surprising that businesses that bank on surveillance would suffer when PETs work well. However, authorities and academics have also cast doubt on Apple and Google's privacy enhancement promises. ATT has not significantly reduced tracking libraries in iOS apps, and developers circumvent ATT with other tracking mechanisms \cite{kollnigGoodbyeTrackingImpact2022a}. PS increases first-party tracking and tracking collaborations between website providers \cite{commissionnationaledelinformatiqueetdeslibertesPubliciteLigneCNIL2024, beuginInterestdisclosingMechanismsAdvertising2024, longEvaluatingGooglesProtected2024}, and its governance structure does not assure that it actually fulfils its privacy promises \cite[][pp. 17-18, 31-32]{competitionandmarketsauthorityCMAQ120242024}. Using PETs to dig deeper into clients could also realise a \textquote{perverse outcome where companies try to use more sensitive data than before as part of their business models} \cite[][p. 46]{vealeConfidentialityWashingOnline2023} since data remains decentralised and confidential -- albeit in the presence of a single party with great control over the underlying infrastructure. These companies narrow privacy to confidentiality \cite{vealeConfidentialityWashingOnline2023, vealeRightsThoseWhon.d.}, data anonymisation, limiting access to personal information, and preventing third-party tracking \cite{mcguiganPrivateAttributesMeanings2023}. Some argue that by framing privacy narrowly to advance their competitive positions, these companies \textquote{sanitiz[e] surveillance} \cite[][p. 9]{mcguiganPartyCynicalResignation2023} and distract from power asymmetries \cite{mcguiganPartyCynicalResignation2023} and \textquote{power dynamics congealed within adtech infrastructures} \cite[][p. 18]{mcguiganPrivateAttributesMeanings2023}. \\

\noindent \textbf{Different views on and types of power.} In all three instances, PETs rely on complex computing and on-device processing. This puts Google and Apple in unique positions to deploy them, granted their control over users' devices, operating systems and browsers \cite{vealeConfidentialityWashingOnline2023}. The analyses provide different perspectives on how PETs enable power. The DP3T analysis exemplifies a \textit{sovereignty view}, showing how Google and Apple asserted themselves as arbiters of privacy and key decision-makers in public health vis-à-vis researchers and governments \cite{troncosoDeployingDecentralizedPrivacypreserving2022a}. The Find My and advertising instances raise \textit{competition} concerns: they underline how these companies gain market persistence or advantage, by using PETs to turn personal devices or browsers into infrastructures for delivering services to third parties. All of this happens while these companies reframe privacy narrowly.
	
The \textit{production view} in this paper empirically studies how such moves impact third parties' production. Beyond scrutinising dependencies and interfaces between manufacturers and tech companies, or even how well the PETs work, we shift to studying: how manufacturers' production processes change in adopting their infrastructures; what role PETs play; and what power dynamics arise. 

\section{Methods and case description}\label{sec_methods_and_case_description}
\subsection{Methods} \label{sec_methods_and_case_description_methods}
We perform a case study \cite{yinCaseStudyResearch2017, swedbergExploratoryResearch2020} into Sidewalk with an empirical \textquote{no theory first} approach. This means we stay as close as possible to the case, given the nascence of relevant literature and the complexity of the instances described therein. We iteratively combine three methods: a grey literature review, technical documentation analysis, and elite interviewing. All methods inform all (sub)questions. 

\textbf{Grey literature review.} With the grey literature review, we aim to (1) understand what Sidewalk promises to users and manufacturers, as well as how manufacturers use it; (2) identify what privacy and security concerns civil society, researchers and (tech) journalists raised vis-à-vis Sidewalk; and (3) add to our broader assessment of how Amazon governs Sidewalk. For (1), we studied Amazon's Sidewalk documentation. In lieu of an Amazon-maintained list of adopters, we thoroughly Googled and searched through all Sidewalk-related publications from Amazon, tech news outlets, and manufacturers to identify third-party manufacturers. We found sixteen organisations; see Table \ref{tab:adopters} in Appendix \ref{appendix_methods}. that have adopted Sidewalk, and their offerings. For (2), we performed a Google search for \emph{``Amazon Sidewalk'' AND privacy}. We supplemented this with a Google Scholar search for \textquote{Amazon Sidewalk}, which yielded few results mostly mentioning Sidewalk in passing. For (3), we analysed Sidewalk and AWS developer documentation and policies \cite{amazonTermsAgreements2023}, and how these changed over time \cite{internetarchiveWaybackMachinen.d.}. For contextualisation, we also briefly studied broader offerings for connectivity (e.g., Matter).

\textbf{Technical documentation analysis.} To understand Sidewalk's framing of privacy risks and safeguards, and to identify what Sidewalk and AWS promise to manufacturers in terms of improving the production of IoT devices, we analysed the technical documentation of Sidewalk and associated AWS IoT services. This helped us understand the workings and capabilities of Sidewalk; its integration with Amazon's cloud; how Amazon processes and secures Sidewalk data; and the requirements that manufacturers' production processes and endpoints must fulfil.	We checked our understanding with four experts in the field of networking and PETs. 

\textbf{Elite interviewing.} Given our focus on production and the context of the case, we adopted an elite interviewing approach to interview manufacturers \cite{latourScienceActionHow1987, loflandAnalyzingSocialSettings1984, solarinoChallengesBestpracticeRecommendations2021, undheimGettingConnectedHow2003, ostranderSurelyYouRe1993, jorgensenParticipantObservationMethodology1989}. The goal is to explore their motivation for and experience with adopting Sidewalk. We asked them how integrating Sidewalk impacts the production of their devices, software and services. The aforementioned materials served as preparation. We identified sixteen Sidewalk-adopting manufacturers and invited 94 C-level executives and department heads through LinkedIn, assuming they possess comprehensive experience with both the business and technical facets of Sidewalk adoption. Eight ultimately participated. We refer to them as \icite{A1}, \icite{A2}, ... \icite{A8}, with the \emph{A} abbreviating \textquote{adopter}. We interviewed one non-adopter \icite{N1}; a C-level executive who is a prominent LoRaWAN industry figure. Respondents signed an informed consent form that informed them they are free to skip questions or withdraw from participation at any time. We conducted all interviews remotely between November 2023 and January 2024, and shared the transcripts with participants, to allow them to redact or edit any statement. Nobody used this opportunity. Our institutional review board approved our overall approach. For analysis, we iterated through two coding cycles, following the manual of \citet{saldanaCodingManualQualitative2021} and using ATLAS.ti 23 \cite{atlas.tiATLASTiDesktop2023}. Appendix \ref{appendix_methods} provides the interview questions and further details about participant recruitment, the coding scheme, and the interview questions.

\subsection{The Sidewalk architecture} \label{sidewalk_architecture}
\begin{figure*}[ht]
	\centering
	\includegraphics[width=0.8\linewidth]{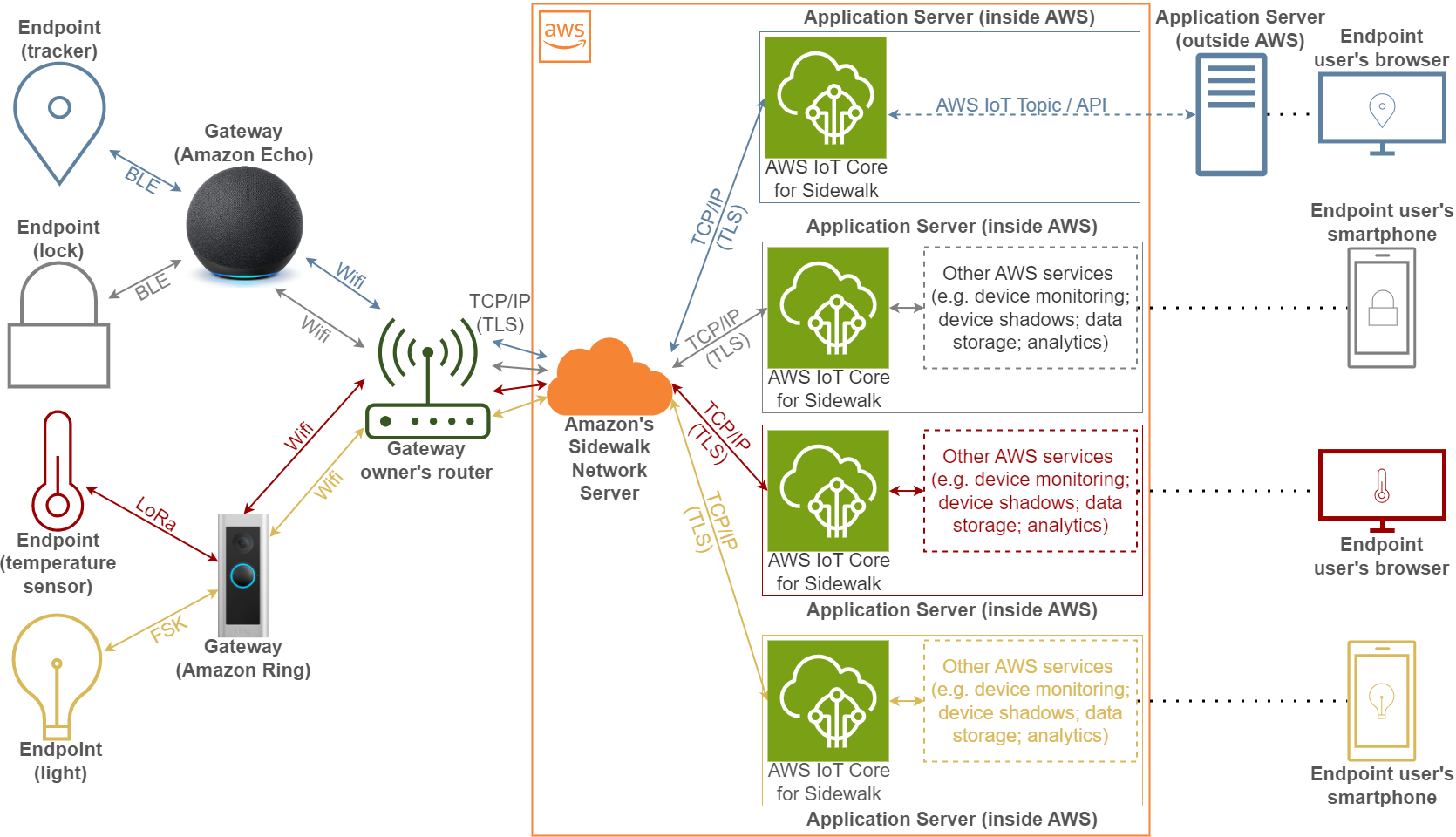}
	\caption[Example Sidewalk architecture]{Example Sidewalk architecture, based on [\citealp[][pp. 14, 42]{amazontechnologiesAmazonSidewalkSpecification2024b};		\citealp[][]{amazonwebservicesWhatAWSIoTn.d.a}]. Echo and Ring images reproduced from \cite{ringVideoDoorbellPron.d., amazonAmazonDevicesEchon.d.}.}
	\Description{Example Sidewalk architecture}
	\label{fig:architecture}
\end{figure*}

Figure \ref{fig:architecture} visualises a hypothetical smart home Sidewalk architecture, where one gateway owner provides connectivity to four endpoints, potentially owned by others. 
We now walk through an example of an endpoint sending data uplink (i.e. to an application server).

Sidewalk-compatible IoT devices are called \textquote{endpoints} and can be produced by Amazon and third parties (\textquote{adopters} or \textquote{manufacturers}). Endpoints contain chips from Amazon-approved silicon providers. To provide endpoints with connectivity, Amazon lets certain editions of their Echo (smart speakers and displays) and Ring (smart cameras, lighting, doorbells, and alarms) devices in the US \textquote{share} a portion of their own wifi bandwidth with nearby endpoints \cite{amazonAmazonSidewalkn.d.}. These devices are called \textquote{bridges} or \textquote{gateways} and can also act as endpoints. The endpoint owner may be someone else than the gateway owner, benefiting from Sidewalk as a \textquote{crowdsourced} network \cite{amazonAmazonSidewalkPrivacy2023}.

Endpoints communicate with gateways using Bluetooth Low Energy (BLE), Frequency-Shift Keying (FSK), or a proprietary version of LoRa \cite[][p. 140]{amazontechnologiesAmazonSidewalkSpecification2024b}. These are intended for short, medium, and long ranges, respectively; with lower data rates for longer distances \cite[p. 11]{amazontechnologiesAmazonSidewalkSpecification2024b}. Not all endpoints, nor all gateways support all protocols \cite{amazonAmazonSidewalkPrivacy2023, amazonAmazonSidewalkGatewaysn.d.}. Gateways check whether endpoint packets comply with the protocol format specifications and that the endpoint is not blocked from Sidewalk \cite{amazonAmazonSidewalkPrivacy2023}. 

Amazon's \textquote{Sidewalk Network Server}, part of its \textquote{cloud} \cite[][p. 15]{amazontechnologiesAmazonSidewalkSpecification2024b}, routes traffic from the gateway to the application server. It inspects all packages; routes them; maintains time synchronisation of the network; and authenticates devices and application servers, verifying that Amazon has not blocked them from Sidewalk \cite{amazonAmazonSidewalkPrivacy2023}.

Last is the manufacturer's \textquote{application server}, that endpoint users interact with through the endpoint's accompanying smartphone or browser interface, e.g. to consult a tracker's location. This latter interaction is outside the Sidewalk scope \cite{amazontechnologiesAmazonSidewalkSpecification2024b}. The server comprises at least AWS IoT Core for Sidewalk and potentially other AWS services. As such, functionalities of AWS include visualising metrics; inspecting, accessing, and updating devices; anomaly detection; and managing digital \textquote{device shadows} \cite{amazonwebservicesWhatAWSIoTn.d., amazonAmazonSidewalkManufacturing2023, amazonwebservicesAWSIoTDevicen.d., amazonwebservicesManageConnectedAWSn.d.}. Manufacturers that want to manage their devices from outside AWS, must interface with IoT Core through a \textquote{Topic} or API \cite[][p. 42]{amazontechnologiesAmazonSidewalkSpecification2024b}. Seven out of eight interviewees picked the former option, thus we only pictured an application server outside AWS for one manufacturer.

With Sidewalk, AWS has a dual role: Amazon uses it to manage the Sidewalk network (i.e. producing connectivity as a service to manufacturers), and manufacturers use it to manage their devices and produce services for their customers. Therefore, we refer to AWS as a software \emph{production environment}.

\subsection{How Sidewalk came to be} \label{sidewalk_history}
Amazon revealed the Sidewalk project during a hardware conference in September 2019, describing a pilot with 700 gateways in Los Angeles \cite{amazonAmazonDevicesEvent2019}. A press release followed soon after, detailing that certain Ring cameras and lights could share their network connectivity with each other \cite{amazonIntroducingAmazonSidewalk2019}. During the pilot, connectivity was restricted to devices registered under the same user account.

In September 2020, Amazon announced that Sidewalk would be launching as a crowdsourced network later that year, and named the first third-party adopters \cite{amazonAmazonSidewalkNew2020}. Simultaneously, they published the first version of the Privacy and Security Whitepaper \cite{amazonAmazonSidewalkPrivacy2020} (further: Whitepaper), that the announcement also references. This is Amazon's first mention of \textquote{privacy} in the context of Sidewalk -- presumably as a response to publications voicing privacy concerns following the 2019 announcements \cite[e.g.][]{warzelOpinionAmazonWants2019, oremusAmazonJustBecame2019, romanoAmazonRollsOut2019, chararaAmazonAppleAre2019a}. 

In May 2021, Amazon announced a further roll-out and more third-party adopters \cite{amazonEchoTileLevel2021}. In June, Amazon transformed Echo and Ring devices of US-based consumers into gateways, through an opt-out OTA update \cite{vaasAmazonSidewalkPoised2021a, callasUnderstandingAmazonSidewalk2021a}. An email notified Echo owners a month before this launch \cite{moorheadAmazonSidewalkFocuses2021}, and Echo and Ring owners received an in-app notification seven days beforehand \cite{vaasAmazonSidewalkPoised2021a}; a narrow window to opt out \cite{vaasAmazonSidewalkPoised2021a, goodinAmazonDevicesWill2021}. In September, Amazon published interviews with executives from two adopters, emphasising that \textquote{Sidewalk's potential begins with privacy protection} \cite{amazonCareBandLife360Tap2021}; presumably again in response to privacy concerns \cite[e.g.][]{nieldHowAmazonSidewalk2021a, callasUnderstandingAmazonSidewalk2021a}. 

\begin{figure}[ht]
	\centering
	\includegraphics[width=0.8\linewidth]{"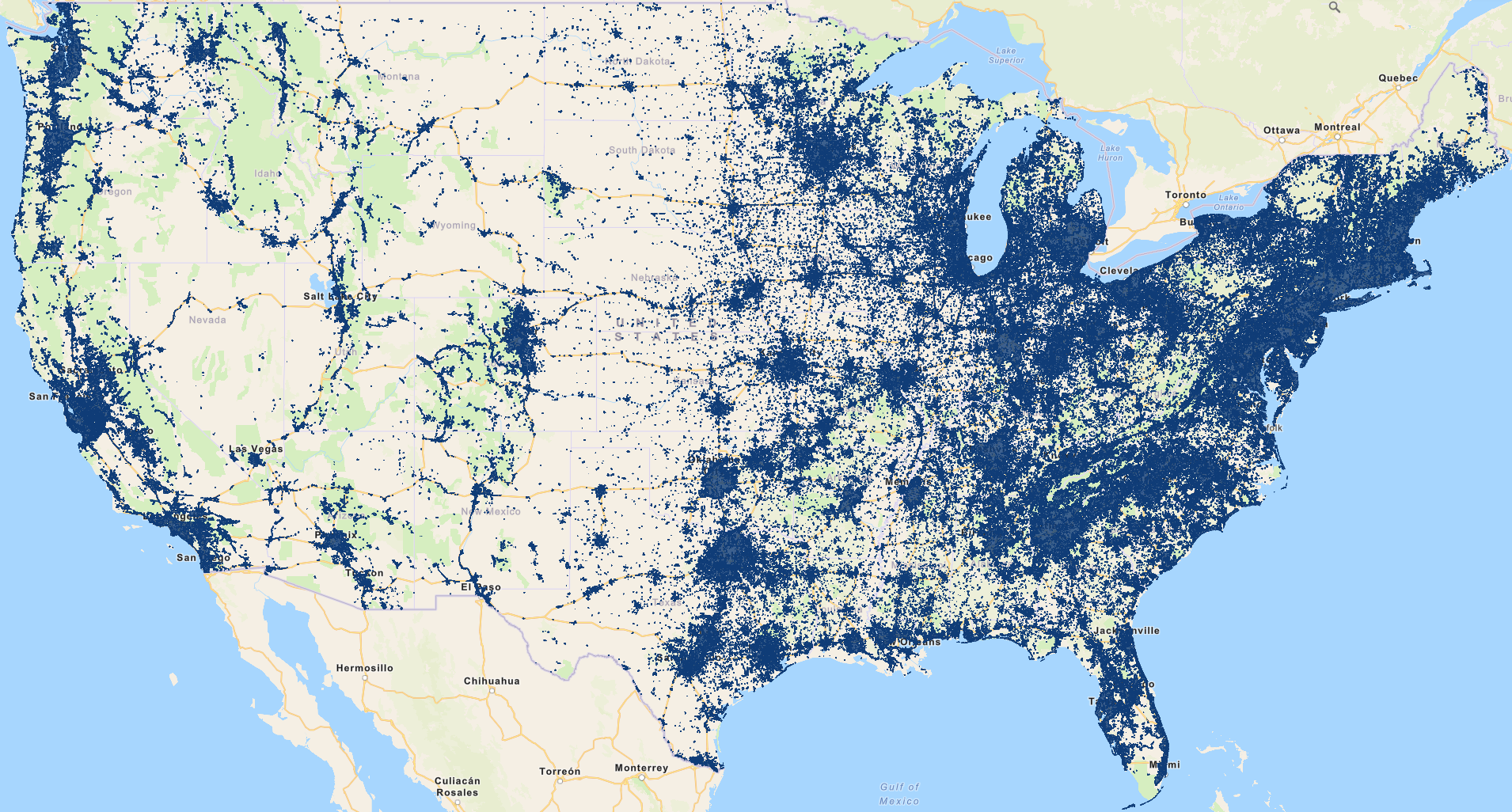"}
	\caption[Sidewalk coverage]{Sidewalk coverage in the contiguous US on November 29th 2024 \cite{amazonSidewalkCoverageMaps2024}. Blue dots indicate coverage.}
	\Description{A screenshot of Sidewalk coverage in the contiguous US on November 29th 2024}
	\label{fig:coverage}
\end{figure}

In March 2023, Amazon claimed to have coverage of over 90\% of the US population (Figure \ref{fig:coverage}) and opened the network for developer testing \cite{amazonAmazonInvitesDevelopers2023}. Shortly after, Amazon updated the Whitepaper to say that Sidewalk was now opt-in for new Echo and Ring devices \cite{amazonAmazonSidewalkPrivacy2022, amazonAmazonSidewalkPrivacy2023}. About 10 months later, an executive proclaimed 95\% coverage \cite{bishopUnfinishedBusinessRing2024}. As of now, the gateway role can be performed by 33 different Echo and Ring models \cite{amazonAmazonSidewalkGatewaysn.d.}, with \textquote{more than 80 million} devices contributing \cite{amazonEmbeddedSoftwareEngineer2024}. 

\subsection{Amazon's marketing of Sidewalk to consumers and manufacturers} \label{sidewalk_marketing}
Amazon promotes different aspects of Sidewalk to consumers (both gateway and endpoint owners; B2C) and manufacturers (B2B) (see Figure \ref{fig:marketing} in Appendix \ref{appendix_marketing_visuals}). 
For consumers, Amazon emphasises community benefits, additional device coverage, security, privacy, and bandwidth constraints. Only the first two advantages demonstrate Sidewalk's functional value; the latter three serve as reassurances. 

On the Sidewalk homepage, Amazon positions Sidewalk as \textquote{a shared network that ... can unlock unique benefits for your device, support other Sidewalk devices in your community, and even locate pets or lost items} \cite{amazonAmazonSidewalkn.d.} (Figure \ref{fig:marketing} \textit{d}). Sidewalk's core proposition is to make IoT devices of all kinds \textquote{work better at home and beyond the front door} \cite{amazonAmazonSidewalkn.d.} by connecting them to the cloud \cite{amazonAmazonSidewalkNew2020, amazonAmazonSidewalkPrivacy2023}.

Gateway owners' participation is central to Sidewalk's coverage. Accordingly, Amazon frames Sidewalk as a \textquote{community benefit}-generating service: a B2C service enabling other B2C services (e.g. Figure \ref{fig:marketing} \textit{b} and \textit{d}). Gateways \textquote{share a small portion of your internet bandwidth which is pooled together to provide these services to you and your neighbors. And when more neighbors participate, the network becomes even stronger} \cite{amazonAmazonSidewalkn.d.}. Gateway owners that opt out \textquote{will no longer contribute [their] internet bandwidth to support community extended coverage benefits such as locating pets and valuables} \cite[][p. 15]{amazonAmazonSidewalkPrivacy2023} \cite{amazonAmazonSidewalkn.d.}. Similarly, one Amazon executive said that Sidewalk \textquote{is best described as a community network} \cite[][19:51]{bishopUnfinishedBusinessRing2024} and another likened it to \textquote{his native village in Southern Spain, where residents make their own soap} and share it with their neighbours, because \textquote{people feel good sharing} \cite{amazonCareBandLife360Tap2021}. Gateway owners allegedly need not worry about helping their neighbours: Amazon warrants Sidewalk's security (Figure \ref{fig:marketing} \textit{a} and \textit{d}; discussed in §\ref{information_flows_and_user_privacy}) and caps gateways' bandwidth usage for Sidewalk at 500 MB per month per customer, at a bandwidth of at most 80 Kbps \cite{amazonAmazonSidewalkPrivacy2023} (Figure \ref{fig:marketing} \textit{c}). \\

\noindent Towards manufacturers, Amazon highlights opportunities for revenue generation and operational control over endpoints out in the world. Amazon paints Sidewalk as \textquote{a secure, free-to-connect, long-range and low-power shared community network designed to provide connectivity for billions of devices} [\citealp[][]{amazonAmazonSidewalkTestn.d.}; similarly in \citealp[][]{amazonAmazonInvitesDevelopers2023} and in \citealp[][p. 11]{amazontechnologiesAmazonSidewalkSpecification2024b}] while also promoting Sidewalk as \textquote{low cost} \cite[][p. 11]{amazontechnologiesAmazonSidewalkSpecification2024b}. More concretely, Amazon's promise is twofold. First are market opportunities: Sidewalk enables manufacturers \textquote{to create and bring to market all types of consumer, enterprise, and public sector smart and connected devices and services} [\citealp[][p. 2]{amazonAmazonSidewalkPrivacy2023}; 
\citealp[][p. 45]{amazontechnologiesAmazonSidewalkSpecification2024b}] Indeed, the sixteen adopters operate across various (not mutually exclusive) domains: logistics (five companies), in-home care (one company), utilities and industry (five companies), and building management (eight companies). Fifteen manufacturers design endpoints specifically for business or governmental users (further referred to as B2B). Nine of these simultaneously sell to consumer users (B2C); although they generally focus on B2B markets. The overwhelming majority of third-party manufacturers thus targets organisations as endpoint buyers. Also, endpoints may be close to homes (e.g. smart lights or motion detectors) or more remote (e.g. asset trackers) \cite{amazonAmazonSidewalkNew2020}.

In making this promise, Amazon depicts Sidewalk as \textquote{a \textquote{pipeline} that moves data back and forth between an Endpoint and its respective Application Server} \cite[][p. 13]{amazonAmazonSidewalkPrivacy2023} (see also Figure \ref{fig:marketing} \textit{c}), and a transport layer: \textquote{{[T]}hey've been very clear [to me] that Sidewalk can do a lot of things: it's a transport layer and you can embed whatever you want into the payload. ... They said, \textquote{remember, Sidewalk is merely a transport shell. What you put into the payload, that's totally your prerogative}} \icite{A6}. 

Second are activities core to a more agile mode of (software) production, native to the AWS cloud environment. Endpoints can only send data to and receive data from an application server that is hosted inside AWS \cite{amazontechnologiesAmazonSidewalkSpecification2024b}. The combination of AWS and Sidewalk connectivity promises manufacturers improved control over devices, and therefore improved development and maintenance of services. Having the data and devices already in AWS lets manufacturers easily connect them to AWS' variety of IoT services that go well beyond offering flexible storage and compute \cite[see e.g.][]{amazonwebservicesManageConnectedAWSn.d.}. Sidewalk's cloud connectivity allegedly contributes to device security (e.g. by offering security-oriented cloud services \cite{amazonwebservicesAWSIoTDevicen.d.} and enabling patching), and also to \textquote{extending their lifecycle and capabilities, cost-efficiency in maintenance from remote operation, and [enabling] rapid prototyping} \cite[][p. 173]{amazontechnologiesAmazonSidewalkSpecification2024b}. To illustrate: with \textquote{Sidewalk Bulk Data Transfer}, manufacturers can send entire firmware updates and files to endpoints \cite[][p. 173]{amazontechnologiesAmazonSidewalkSpecification2024b}. In that sense, Sidewalk and AWS can make manufacturers' production more agile \cite{gursesPrivacyAgileTurn2018}.
 
\section{Information flows and PETs in Sidewalk} \label{sec_information_flows_and_PETs_in_sidewalk}
Sidewalk uses two classical PETs: end-to-end encryption (with three layers, shielding both payloads and metadata) and obfuscation (of endpoint and gateway credentials, and routing information) [\citealp[][]{amazonAmazonSidewalkPrivacy2023}; \citealp[][]{amazonComponentsAmazonSidewalk2023}; \citealp[][chapter 4]{amazontechnologiesAmazonSidewalkSpecification2024b}]. A short elaboration of each is given in §\ref{sidewalk_PETs}. PETs in Sidewalk serve \textquote{to secure data traveling on Sidewalk and to keep customers safe and in control} \cite{amazonAmazonSidewalkn.d.} and limit \textquote{the collection and storage of customer information} \cite[][p. 3]{amazonAmazonSidewalkPrivacy2023}. The design principles are that gateway owners \textquote{do not receive any information about devices owned by others connected to Sidewalk} \cite{amazonAmazonSidewalkn.d.}; that \textquote{{[i]}nformation [endpoint] customers would deem sensitive, like the contents of a packet sent over the Sidewalk network}, is not visible to Amazon or gateway owners \cite[][p. 4]{amazonAmazonSidewalkPrivacy2023}; and that \textquote{tracking of devices and associating a device to a specific user} by Amazon, gateway owners, and eavesdroppers is prevented \cite[][p. 4]{amazonAmazonSidewalkPrivacy2023}. A traditional privacy analysis would look to see whether the implementation of these PETs leaves privacy risks unaddressed. We briefly do so (§\ref{information_flows_and_user_privacy}), but then take a step back to deliberate the role of PETs in \textquote{flipping} consumer devices for extending Amazon's computational infrastructure. 

\subsection{Information flows and user privacy} \label{information_flows_and_user_privacy}
Sidewalk limits some information flows with PETs while, simultaneously, casting a vast crowdsourced network that enables \emph{additional} information flows. For instance, Sidewalk obscures to gateway owners whether their device provides an asset tracker with connectivity; but this connectivity enables information to flow from the tracker to the owner, its manufacturer, and Amazon. These new information flows are irrespective of the original communications between the device and cloud necessary for delivering services to the customers that bought them. Sidewalk enables other users to remotely manage their devices; manufacturers to collect telemetry and deploy OTA updates; and Amazon to optimise its IoT device and cloud service offerings. These capabilities are enabled both by Sidewalk's connectivity and the range of AWS services. 

Considering these new information flows, we identify four types of unresolved privacy concerns: (1) increased surveillance of consumers by other consumers and law enforcement; (2) Amazon's and manufacturers' increased ability to monitor devices and user interactions; (3) further blurring of boundaries between private and public life; and (4) undermining of owners' control over their personal devices. We explain these concerns in Appendix \ref{appendix_information_flows_and_PETs}. Herein, we note that the proprietary design and a lack of transparency of Sidewalk's technical workings further aggravate these concerns; a red flag for privacy and security communities. PETs researchers have repeatedly proposed (partial) solutions to such problems \cite[e.g.][]{jakariaConnectingDotsTracing2024, sunSoKSecureHumancentered2024, wangPrivacyGuardExploringHidden2025, sunRevealingHiddenIoT2025}, but these are not part of Sidewalk's PETs. We conclude that Amazon takes a narrow view on the privacy concerns that may arise from Sidewalk. 

\subsection{The role of PETs in flipping consumer devices} \label{PETs_role_in_flipping_consumer_devices}
Amazon came with multiple responses to the privacy pushback that Sidewalk generated over the years. Most obvious is publishing and repeatedly referencing the Whitepaper that outlines Sidewalk's privacy, security, and bandwidth restriction measures (§\ref{sidewalk_history}). Further, Amazon obfuscates the predominant B2B nature of third party devices by repeatedly pointing to B2C applications (§\ref{sidewalk_marketing}). Additionally, an interviewee recounts that Amazon strategically released stories of adopters' products, to demonstrate Sidewalk's value and distract from the backlash. Amazon used them \textquote{as a pawn basically in the game, of showing that there's value to this network}. Amazon also called another participant \textquote{a very strong opportunity for Sidewalk to show off what it can do}. Together, Amazon uses these angles to craft a narrative that their transformation of personal devices benefits gateway owners: they can help their community without having to worry about bandwidth and privacy.

This strategy suggests that Amazon had to address privacy concerns to justify \textquote{flipping} consumers' Echo and Ring devices to include gateway functionality; leading Amazon to widely advertise the PETs they put in place. \icite{A2} confirms this reading: \textquote{it's very obvious if you look at the timeline of when things were released and how they were released and the emphasis on things ... They were very aware of the fact that people could have an adverse reaction}. Once Echo and Ring devices are \textquote{flipped}, information and control flow between other endpoints and their users, manufacturers, and Amazon. With consumer devices flipped to gateways, Amazon \textit{extends their AWS production environment} with a network covering over 90\% of the US population, while sparing the company the costly endeavour of building out and maintaining a new physical connectivity infrastructure. Echo and Ring owners buy a device, deploy it across the country, provide it with wifi and electricity, and troubleshoot it; and ISPs move the packets between gateways and Amazon's cloud. In exchange, gateway owners are offered narrow privacy guarantees and a narrative of altruism -- if they discover their participation at all.\footnote{A survey of users' perceptions of Sidewalk and satisfaction with its proclaimed benefits would be interesting but goes beyond the research question of this paper.} 

\subsection{Information flows and manufacturers' confidentiality} \label{information_flows_and_manufacturer_confidentiality}
\noindent The new information flows in Sidewalk benefit IoT manufacturers but, as our documentation review and interviews showed, also expose business-sensitive information to Amazon. Amazon has insight into a swath of telemetry (see Table \ref{tab:endpoint_data_metadata_telemetry} in §\ref{endpoint_telemetry_visibility}). This data lets Amazon monitor the integrity and reliability of Sidewalk \icite{A2; A6}: \textquote{they need metrics to see if things actually work} \icite{A2}.

But with this telemetry, Amazon can also \emph{learn the demand for endpoints} of each manufacturer and thus application domain. Two interviewees said they expect, and two confidently asserted that Amazon is using their vantage point for this purpose. \icite{A1; A4} brought up that Amazon has a history of doing so on their Marketplace, recurrently utilising data of third-party sellers to replicate high-demand products under their own brands \cite{mattioliAmazonScoopedData2020, kalraAmazonCopiedProducts2021, europeancommissionAntitrustCommissionAccepts2022}. Two other respondents think this benefit is negligible.

Only one interviewee was concerned about this premise, \textquote{but in the end there's not a whole lot [they] can do about it, you know, aside from government action or lawsuits}. Five were not concerned, thinking Amazon will not enter their line of business \icite{A3; A5} or believing in their own company's capabilities \icite{A2}, intellectual property and patents \icite{A3}, and established brand name \icite{A3} to outcompete them (\textquote{Bring them on!} \icite{A2}). Further, multiple participants believe that Amazon's Sidewalk endeavour, and them potentially selling similar competing endpoints, helps educate and grow their markets. \icite{A2; A7; N1} pursue short-term revenue gain while acknowledging long-term risks for their competitiveness vis-à-vis Amazon, in what one interviewee called \textquote{coopetition}. \\

\noindent Beyond demand, telemetry might inform Amazon about endpoint \textquote{power usage, connectivity usage, and data transfer} patterns \icite{A8}, and \textquote{help them if they were gonna do things like [design] their battery model for their [device] and figure out how many times someone [triggers the device] or something like that} \icite{A1}. Three interviewees supposed that the encryption of payload data inhibits such learning (that Table \ref{tab:endpoint_data_metadata_telemetry} disproves), or that this knowledge is not so valuable for endpoint development that it would warrant \textquote{stealing information}. \\

\noindent Finally, Amazon can improve AWS for IoT manufacturers at large, with its insight into endpoint behaviour and how manufacturers manage them in AWS. As \icite{A4} put it: \textquote{To credit Amazon, with AWS in particular, ... they also did a great job of seeing what customers were doing with their products and introducing new services that better met those needs. So they're very good at that. To what extent are they doing that with Sidewalk? I honestly don't know. But I'm sure they are.} Optimising AWS as a software production environment aligns with \icite{A5}'s conception of the AWS business model: \textquote{the business model is \textquote{data has gravity}. So they want the data coming to AWS. ... That's where they make their money}. This notion posits that collecting large amounts of data in one environment increases the odds of attracting other data and services therein \cite{mccroryDataGravityClouds2010}. 

\subsection{Instrumentalisation of PETs to extend Amazon's computational infrastructure} \label{PETs_to_extend_CI}
\noindent When we dig deeper, we see that while Sidewalk implements PETs to provide some privacy, the story is more complex. Amazon reduces privacy to the confidentiality of certain user data, that it can address with PETs. With these concerns out of the way, Sidewalk flips consumer devices to enable new information flows. The consumer devices-cum-gateways extend Amazon's computational infrastructure; that now offers cloud services and connectivity as a production environment to IoT manufacturers. Meanwhile, Sidewalk's PETs expose an area that covers 90\% of the US population to novel privacy concerns. The PETs also do not shield manufacturers' business-sensitive information from Amazon. Interviewees responded to this with resignation, nonchalance, or a bet on coopetition; with very few voicing concerns. 

Amazon, we conclude, instrumentalises PETs in Sidewalk to transform personal devices to gateways and pit themselves in a powerful vantage position vis-à-vis manufacturers. This is similar to \citeauthor{vealeConfidentialityWashingOnline2023}'s [\citealp{vealeConfidentialityWashingOnline2023}] conclusion that tech companies use PETs not only to encrypt communication, but also to enable novel cross-party functionality, in our case through new information flows.
 
\section{PETs transforming production} \label{sec_PETs_transforming_production}
Moving beyond information flows, we investigate the actual engineering and governance of PETs in Sidewalk. Amazon not only specifies privacy, and associated security requirements, but takes an active role in defining their implementation. We focus on how implementation arrangements change production of IoT manufacturers as their devices become accessories of Amazon's AWS.

\subsection{Implementation of encryption in endpoints} \label{implementing_encryption}
Amazon's implementation of PETs in Sidewalk requires manufacturers to embed encryption keys and certificates in endpoints; a process that interviewees called \textquote{keying}. Factories must use a \textquote{YubiHSM} hardware security module (HSM) \cite{yubicoYubiHSM2V232n.d.} programmed to sign the device certificate with the private key of the manufacturer, so that only the endpoint and manufacturer's application server can decrypt each other's packets. The manufacturer must purchase the HSM and send it to Amazon for \textquote{factory support} \cite{amazonAmazonSidewalkManufacturing2023}. This entails Amazon programming a \textquote{device attestation key} and the \textquote{Sidewalk certificate chain} onto the HSM \cite{amazonAmazonSidewalkManufacturing2023}. Thereafter, the manufacturer sends the HSM to the factory, that connects it to a computer on the factory line \cite{amazonHowManufactureProduce2023}. The factory must also share logs of manufactured devices (including unique persistent identifiers) with Amazon, for routing and authentication by the Sidewalk Network Server.

In the interviews, implementing the keying workflow surfaced as \textquote{affect[ing] the actual \emph{physical} production, quite severely. Because Amazon has stringent requirements on security for production. They have very high standards on how they actually get keys into devices. ...for us, it was a minor change. But it was still a minor change. It was not ... just plug and play} \icite{A2}. For instance, it is not standard for factories to use computers in their actual manufacturing process \icite{A1}. For manufacturers that are relatively small or have less experience, this is \textquote{a big change. It's a big step up in security. ... if you don't have these kind of systems from before when you do production, then it's a significant change. ...it will take some while to get the production of this up} \icite{A2}. Especially manufacturers of less sophisticated devices might struggle: \textquote{it's definitely a hurdle to get over, ... especially if it's something simple like a light or something, like \textquote{I just want to make a light and ... why do I have to figure out how to provision all this stuff in order to make it work?}} \icite{A1}. The keying requirement showcases Amazon's ability to go beyond specification to stipulating workflows that change how IoT manufacturers organize their physical production process.

\subsection{Transmission and hardware requirements} \label{transmission_and_hardware_requirements}
Amazon does not merely specify capabilities that hardware must meet to implement their PETs, but limits manufacturers to four approved partner development kits \cite{amazonWorksAmazonSidewalk2023b, amazonQualifiedDevelopmentKits2023a}. According to \icite{A7}, \textquote{only certain processors [are] allowed to operate on Sidewalk because they have to have that encryption zone built into it. You have to have a certain amount of memory. You have to handle data in certain ways}. Additionally, endpoints must use the AWS-developed real-time operating system FreeRTOS and comply with data rate, volume, synchronism, and frequency requirements in radio transmissions. 

That Amazon goes beyond hardware requirements to using four \textit{qualified} silicon providers has varied affects on IoT manufacturers. The Sidewalk stack is hard to implement: \textquote{especially in the very beginning, the efforts to make it work were pretty complex}. Four interviewees noted that the stack requires relatively powerful and therefore costly hardware, e.g. in terms of memory. The cost is especially problematic in the B2C domain, where price-sensitive consumers lead manufacturers to minimise device costs \icite{A1; A6; A7; N1}. The high firmware footprint means that some manufacturers could not fit additional protocols in the same device. Another B2C-focused participant indicated that their current endpoints could not accommodate the firmware and saw insufficient value in Sidewalk to use more powerful hardware. In response, Amazon created special software for them: \textquote{the Sidewalk implementation with [our devices] on both sides is quite custom}; \textquote{We have custom APIs, they have built custom firmware for [some gateways] around this. Yeah, it's a very [our company]-specific implementation}. Consequently, their endpoints do not use the Sidewalk authentication and security scheme. Amazon apparently permitted this security lapse for the sake of onboarding this particular organisation. They did not extend the privilege to other partners as the pool of adopters grew, underlining Amazon's sole decision power over \textit{qualified} implementations. Finally, one manufacturer said that the FreeRTOS requirement came after their admission to the Sidewalk program. It forced them to move from coding for bare metal, to coding for different such OSes and ultimately FreeRTOS (stewarded by AWS); demanding a fundamental change in development that Amazon provided little transparency around.

\subsection{Organisational governance and certification} \label{organisational_governance_and_certification}
Contrary to similar protocols such as Bluetooth, Matter, and LoRaWAN, Sidewalk is not governed by a standards body but solely by Amazon. Amazon is also the only party operating the network's backbone, i.e. the Network Server and gateways. 

Amazon effectuates their implementation requirements through an \textit{ex-ante} qualification process and its encryption key infrastructure. To qualify, manufacturers must request a \textquote{Sidewalk360} account, inform Amazon about their organisation and product, and obtain \textquote{development keys} to connect the prototype to Sidewalk during testing; have a test facility prove that prototypes pass the Sidewalk test cases; and pay a qualification fee (currently zero) \cite{amazonWorksAmazonSidewalk2023b, amazonAmazonSidewalkQualification2023a, amazontechnologiesAmazonSidewalkTest2023}. Only after approval may manufacturers advertise Sidewalk compatibility and may their devices access the network.

Complying with the Sidewalk policies is a continuous process and prone to change~\cite{amazonWorksAmazonSidewalk2023a}. Endpoints must be reliable and contribute to \textquote{an overall good customer experience}, e.g. not suffer from repeated disruptions or latency \cite{amazonWorksAmazonSidewalk2023a}. Furthermore, manufacturers must provide security updates for endpoints as long as Amazon says so (\textquote{no less than 4 years from the last shipping date of the device} \cite{amazonAmazonSidewalkQualification2023a}); address security vulnerabilities that the manufacturer encounters within a time period that Amazon defines \cite{amazonWorksAmazonSidewalk2023a}, immediately notify Amazon, and \textquote{take all appropriate steps to remedy such vulnerability, including cooperating with [Amazon]} \cite{amazonAmazonSidewalkProgram2023}; and propagate Sidewalk updates -- that can be for \textquote{firmware update, reporting, or debugging purposes} -- to all endpoints, again within a certain period, and share the collected metrics with Amazon \cite{amazonAmazonSidewalkProgram2023}. Ironically, the terms forbid manufacturers to monitor \textquote{the availability, performance, or functionality of any of [Amazon's] products or services} \cite{amazonAmazonSidewalkProgram2023}.

In February 2024, Amazon adjusted the uplink traffic rates and introduced a daily limit, a top-down change in policy [\citealp[][]{amazonHowAmazonSidewalk2023}; \citealp[][p. 18]{amazontechnologiesAmazonSidewalkSpecification2024b}] Amazon reserves the right to re-audit or recertify adopters at later stages, and arranges periodic check-in meetings with adopters to elicit their feedback \icite{A6}. Amazon may retroactively revoke authentication keys from infringing manufacturers and \textquote{if a third party fails to act in good faith} \cite[p. 14]{amazonAmazonSidewalkPrivacy2023}. \\

\noindent Amazon has firmly established themselves as Sidewalk's sole governing party, 
putting the company in a position to unilaterally impose obligations, arbitrarily deviate from their policies, and liberally interpret their broadly-formulated obligations (e.g. the \textquote{good customer experience}); exposing manufacturers to power asymmetries. While interviewees generally considered the qualification process to be reasonable, their experiences show that it is less standardised and runs deeper than the documentation describes. One participant incurred a delay in their prototypes testing: Amazon kept on retesting them because they were not finding anything wrong with it. Furthermore, Amazon performed in-person security inspections of the factory that a participant worked with, although this is not mentioned in the qualification documentation.

Amazon was said to leverage its inspections for commercial reasons: i.e. polishing Sidewalk's image. An early adopter shared Amazon's interest in their finances: \textquote{they audited the company, they audited the objectives of the company ... They had to understand how we were being financed, to determine if we're going to be around for a while, or if we were just a start-up that was going to disappear in 6 months}. Amazon also verified adopters' \textquote{market potential} so that their endpoint could contribute to Sidewalk's marketability (§\ref{PETs_role_in_flipping_consumer_devices}).

Amazon took the liberty to give some manufacturers a favourable treatment, depending on their relations. One respondent's organisation only partially completed the qualification process in order to bring Sidewalk functionality to endpoints later with an OTA update. This is \textquote{not typically how you do it} and a privilege they ascribed to \textquote{the world [being] a small place}, as they were acquainted with someone leading Sidewalk. An early-adopter interviewee had a \textquote{little bit different but ... similar process}, \textquote{because we've been working kind of hand-in-hand with them for a while}. One participant said they would not even undergo qualification, despite Amazon and themselves advertising their Sidewalk adoption. 

With their governance process, Amazon obtains long-lasting agency over how endpoints work and how manufacturers produce and manage them at what rhythm. Meanwhile, the lack of checks and balances means Amazon can hardly be held accountable for favouritism and arbitrariness in enforcement. \\

\noindent Amazon's sole reign also implies that manufacturers and users rely on their dedication to maintain the Network Server and gateways. If Amazon were to pull the plug on either, all endpoints would lose Sidewalk connectivity. Granted repeated lay-offs \cite{ortakalesAmazonLayoffsTimeline2024} and Amazon losing billions of dollars on their devices division between 2017 and 2021 \cite{mattioliAlexaMillionsHouseholds2024}, this poses a realistic risk.

One respondent says Amazon is \textquote{in the works of creating a development kit} to enable manufacturers to have their endpoint double as a gateway, which their company is anticipating and developing towards. When asked why Amazon would not enable this from the get-go -- and considering that we found no other information about the kit, the participant replied: \textquote{I can't say that they've ever given us a true answer ... there's a lot of vagueness. ... I'm wondering if it's because they still wanna hold that control ... I see it happening eventually, I just can't give you a timeline, because they don't give us a timeline}. Manufacturers internalise the risks of such uncertainty: \textquote{if they shared a little bit more, then we could develop a gateway with them. But you know, we're not there}.
	
\subsection{Application server in AWS} \label{AS_in_AWS}
AWS ties all the ends together in the implementation of Sidewalk's PETs. Sidewalk requires that all traffic originates from or is sent to an application server within AWS \cite{amazontechnologiesAmazonSidewalkSpecification2024b}: \textquote{there's no option for the data not going to AWS} \icite{A5}, because \textquote{Sidewalk is decoded and decrypted in Amazon's cloud} \icite{A6}. This decoding and decrypting entails both the en- and decrypting of packets, and the authenticating of devices and application servers \cite{amazonAmazonSidewalkPrivacy2023}. Once manufacturers land on AWS, they can utilise the myriad of cloud services part of AWS' production offerings (see §\ref{sidewalk_architecture}). The operational control that these services yield are important reasons for manufacturers to adopt AWS \icite{A2; A6; A8}. 

While seemingly convenient, Amazon's tight integration of Sidewalk with AWS severely hinders manufacturers to manage endpoints from a non-AWS (cloud) server. If manufacturers (or their business customers) wish to use other non-AWS infrastructure, they must move data around or expose APIs from within AWS to these other servers. \icite{A5; A6; A7} have such APIs that their business customers can interact with. \icite{A5} said that \textquote{if a [customer] wanted to use Azure, then we could just get the data to Azure, that's fine, that takes ... a fraction of a millisecond. ... You gotta pull it out, it'll go into your AWS S3 bucket and go into a place, and then you can get that to wherever that needs to go. ... Moving data around is a pretty solved problem these days.}. \icite{A7} thinks it is \textquote{doable}, but \textquote{not as easy as is said} because \textquote{there's still a copy and paste effort}.

Regardless of whether it is easy, we underline that the data has to be moved around in the first place, as AWS cannot be avoided. This constitutes a duplication effort that entails increased complexity, data integrity challenges, and security risks. Accordingly, Amazon \textquote{funnel[s] people} into using AWS \icite{A7}.

This funnelling has two discursive consequences. First, we cannot refer to Sidewalk as a wireless mesh network (WMN) \cite[as e.g.][do]{songAreAmazonSidewalk2023, callasUnderstandingAmazonSidewalk2021a, lardinoisAmazonSidewalkAdds2023a}. Contrary to a WMN \cite{akyildizWirelessMeshNetworks2009}, endpoints and gateways cannot communicate bidirectionally and locally over Sidewalk. This promotes the model of production where all IoT communications pass over an application server. For instance, one study found that about 65\% of IoT device companion apps use Bluetooth, implying that 35\% has network- or cloud-based communication \cite{schmidtIoTFlowInferringIoT2023}.

Second, the funnelling demonstrates that Sidewalk is not \textquote{just a pipe}, contrary to Amazon's framing (§\ref{sidewalk_marketing}). In net neutrality discussions \cite{thiererAreDumbPipe2005}, connectivity and hosting providers would argue that as a \textquote{dumb pipe} or \textquote{mere conduit} (i.e. linking end users to content providers over the internet, without moderating content) \cite{dediegomartinNetNeutralitySmart2016, jamisonNetNeutralityPolicies2018}, they have reduced liability over content \cite{rendaTelecommunicationsInternetTTIP2015, husovecDigitalServicesAct2023, vangeunsHowHateSpeech2020}. Amazon's appeal to being a pipe implies that they barely intermediate in Sidewalk; granting manufacturers and endpoint users autonomy in deciding what data they send where. On the contrary, Amazon embeds Sidewalk connectivity closely in AWS and its IoT-related services. Sidewalk thus does not provide internet connectivity, but \textit{cloud} connectivity [\citealp[][]{amazonIntroductionAmazonSidewalk2023}; \citealp[][p. 9]{amazontechnologiesAmazonSidewalkSpecification2024b}]. This setup incentivises manufacturers to utilise AWS services for managing their endpoints, allowing Amazon to monetise Sidewalk indirectly. Meanwhile, Amazon obtains the power to shape what endpoints can(not) do, by virtue of both designing Sidewalk and the IoT services in AWS that manufacturers use; going far beyond what a \textquote{pipe} would do.

\subsection{How PETs give Amazon an upper hand in production} \label{PETs_upper_hand_production}
Amazon's marketing of Sidewalk as a crowdsourced network that benefit consumers, obscures the greater movements at play. Sidewalk does not only allow Amazon to extend their computational infrastructure with a crowdsourced network, but also turns third party IoT devices into accessories of their cloud. Manufacturers can benefit from this: they can now utilise cloud-based remote device management functionality, such as communicating with and OTA-updating of resource-constrained endpoints. They can further monitor service use and other telemetry to iterate over services in an agile manner~\cite{gursesPrivacyAgileTurn2018}. As our interviewees highlighted, this brings benefits for security patches, but also for continuously updating service functionality to reflect changes in business objectives and outcomes of the service provider. 

To receive these advantages, however, manufacturers need to accept the interventions Amazon makes into their production as part of their implementation of PETs. Amazon's control over the implementation allows it to intervene in endpoint hardware (e.g. qualified chips), software (e.g. the large and complex software that must be kept up-to-date), operating systems (e.g. the requirement to implement FreeRTOS) and the environments producing and managing them (e.g. the mandatory use of AWS, keying of devices, and organisational governance). Further, by adopting Sidewalk, manufacturers enter a path dependency that aggravates these power asymmetries. Sidewalk demands significant and perpetual technical (hardware, software, cloud) and organisational (staff time, knowledge, and production lines) resources that cannot be spent otherwise. Multiple interviewees indicated this made it hard for them to adopt and retain knowledge about other connectivity protocols. As \icite{A3} aptly summarises: \textquote{it's hard trying to figure out where you put your eggs, where you focus, is a really important question for viability of a business}. Indeed, one interviewee did not yet turn Sidewalk on despite already making their device hardware compatible: \textquote{we would need to prioritize the R\&D bandwidth to actually do the work, versus all other products or projects. So then it becomes a \textquote{portfolio management} kind of thing} \icite{A2}. Finally, Sidewalk's funnelling of manufacturers into AWS raises the bar for them to adopt other cloud services. Thus, once in, shifting away from Sidewalk and AWS becomes challenging; both because manufacturers are distanced further from alternatives and because Amazon's infrastructure and governance mechanisms become so deeply engrained in their production process, that they perpetually reshape and complicate them. This entrenching also exemplifies Amazon's power to reorganise business relations between Amazon, Sidewalk adopters, and the companies they source their hardware components from.
 
\section{Discussion} \label{sec_discussion}
With our study of Sidewalk, we explore whether and how Amazon instrumentalised PETs to entrench their computational infrastructure. Vis-à-vis consumers, we found the introduction of PETs to Sidewalk were key to apprehending pushback against flipping their devices into gateways. With Amazon's narrow definition of privacy and rudimentary use of PETs, Sidewalk suppressed some sensitive information flows, to cast a network with new information flows that Amazon leverages as part of its offering to IoT manufacturers. By exerting sole control over the requirements, design and implementation of PETs in Sidewalk, we showed Amazon entrenches its extended infrastructure in and gains agency over manufacturers' production, putting into place potentially long-lasting power asymmetries. Sidewalk's use of PETs therefore not only has implications for users of gateways and endpoints, but also, if not more, for endpoint manufacturers.

\subsection{PETs in the ``clipping economy''} \label{clipping_economy}
The power to flip consumer devices into an extension of a company's computational infrastructure requires both existing infrastructure and skills in implementing PETs. Amazon's capability to push an OTA update to a critical mass of adequate devices (i.e. with hardware appropriate for transmission over LoRa, Bluetooth, and FSK, and for the PETs) and control over their OS surface as technical preconditions. Another is to convince consumers (and privacy regulators) that use of personal devices for delivering services to third parties does not impede upon the privacy of actors involved. This is a repeating pattern. Apple's crowdsourced finding service; Google and Apple's privacy-preserving advertising environments; and the digital contact tracing framework by the same companies; all come with PETs. In all instances, a company's control over a critical mass of consumer-owned clients -- be it browsers or actual devices -- combined with more or less sophisticated PETs, allows them to leverage those devices into an infrastructure for offering novel services to consumers, businesses, and governments. Moreover, these companies leverage the necessity to control information flows and secure the implementation of PETs to centralise their role in and control of this extension, often to their benefit. 

Using consumer devices to create a closed mesh network for services to third parties, with greater benefits for a provider, is not new. ISPs now and before have added additional closed wifi networks using routers belonging to their customer base. This allows the ISPs to offer a wifi network to customers outside of their homes or to other customers with a dedicated wifi hotspot subscription \cite[e.g.][]{xfinityXfinityWiFiHomen.d., deutschetelekomMobilityOnlineAny2013, fonFonOTEAnnounce2014, britishtelecommunicationsBTWiFin.d.}. What differentiates Sidewalk is that it does not just provide connectivity to consumers, but also to other businesses; i.e. IoT manufacturers and their customers. Moreover, Sidewalk hooks endpoints directly onto Amazon's cloud, making seamlessly available to manufacturers their plethora of IoT device management and data processing services in AWS IoT Core and in AWS more broadly. This allows Amazon to indirectly monetise Sidewalk, without charging manufacturers or users to access the network itself. 

More generally, analogous to but different from situations in which service providers \textquote{clip} a part of third-party revenue (e.g. Apple's App Store and Stripe's payment platform taking a percentage of revenue), we consider Sidewalk and these other instances as a new chapter in a wider development that we refer to as the \textquote{clipping economy}. In these instances, tech companies clip some benefits from the \emph{infrastructure} made up of devices in the hands of consumers. With digital contact tracing and finding networks, Google and Apple use smartphones as an infrastructure for delivering sensing services to governments and consumers. In Privacy Sandbox and App Tracking Transparency-adjacent technologies, Google and Apple use browsers and OSes for on-device PETs to facilitate ad delivery for businesses. Sidewalk goes even further: Amazon appropriates a portion of gateway owners' wifi bandwidth and therefore of telecommunication providers' infrastructure.

\subsection{The Privacy Enhancing Technology Paradox} \label{PET_paradox}
Contrary to PETs' general aim to tackle power by inserting strict controls over information flows, Amazon's implementation of PETs gives rise to a two-faceted \emph{Privacy Enhancing Technology Paradox} (\emph{PET Paradox}). First, tech companies can adopt PETs with a narrow interpretation of privacy to flip consumer devices and enable novel flows of information. With this, they may expand their infrastructural power, compromising consumer privacy and the confidentiality of companies adopting their infrastructures, respectively. Consequently, when assessing how well PET implementations address specified privacy concerns, one should look beyond privacy promises and assess other novel risks that may arise.

Second, devices becoming accessories of clouds has advantages for implementing privacy (and security) but also grave disadvantages for the design of PETs. In modern-day service architectures, devices are increasingly managed and programmable accessories. This is a win for privacy, as manufacturers can patch vulnerabilities of devices already in use, or deploy and update decentralised PETs -- in the case of Sidewalk even without needing user intervention. However, control over the mechanisms for remote programmability is also what enables Amazon to \textquote{flip} Echo and Ring devices. In §\ref{sec_PETs_transforming_production}, we saw how Amazon's requirements for device hardware, software, and communications (i.e. to and through AWS) insert Amazon's infrastructural control deeper into these third-party devices. These developments indicate a web of control around devices that erode assumptions PETs researchers can make about devices being a trusted or personal base for decentralised PETs. Ironically, the Sidewalk case demonstrates how Amazon controls the implementation of their PETs to make interventions into manufacturers' devices, cloud, factory lines, and organisational processes. These interventions entrench Amazon's computational infrastructure -- now extended to include consumers' gateways -- in the process. This second facet of the PET Paradox urges us to reflect on the viability of PETs as a few companies control computational infrastructures in the form of clouds and end devices as their accessories. 

\subsection{Betting on rising with giants} \label{rising_with_giants}
The power asymmetries between Amazon and manufacturers raise the question: do manufacturers think Sidewalk brings sufficient functional benefits to risk these dependencies on Amazon and their infrastructure? Multiple participants mentioned that having other business relationships with Amazon (e.g. having \textquote{a major retail relationship} or using AWS) influenced their decision to adopt Sidewalk; even if its requirements or bandwidth limitations mean Sidewalk is \textquote{not a perfect fit for [them]}, as one interviewee said. One participant expressed that \textquote{there’s an aspect of an ongoing partnership with Amazon, which is a huge company}. Another respondent relies on and envisions future integrations with Amazon's logistics and parcel delivery services, seeing Sidewalk as \textquote{a stepping stone towards future development ... It can create a more intimate relationship with Amazon to wanna do future developments for [our company]. I would say that's probably the point.} Indeed, \textquote{it's like befriending the giant, right? ... If we create a rocky relationship with them and shut them out, well, that's going to eliminate a huge portion that we could have for business with them. And we really aren't trying to create that kind of rough rockiness}. But to do this, the interviewee adjusts to Amazon's moves: \textquote{There's always about, you know, who's holding the more power ...? And we just kind of play ball the way they play ball, and hopefully develop an innovation that can drive the growth. That's all}. In sum, Amazon's infrastructural and market power (i.e. being a big player in the cloud, logistics, retail, and home IoT domains) beget more power (i.e. the additional power that Sidewalk gives them).

\section{Limitations} \label{sec_limitations} 
Our work knows multiple methodological limitations. First, interviewing is time-consuming, necessitating a limit on the number of interviews and reducing the generalisability of the findings \cite{alshenqeetiInterviewingDataCollection2014}. Additionally, the population of Sidewalk-adopting manufacturers is limited to sixteen companies. While our eight Sidewalk-adopting interviewees cover half of this group, others might have different stories about Sidewalk's effect on their production. Further, with eight out of nine interviewees having adopted Sidewalk, the interviewee sample is likely positively biased towards Sidewalk. Interviewing more non-adopters to hear more about their rationales would be interesting.

Similarly, the consulted grey literature was frequently biased, inaccessible, and incomplete; complicating analysis. Three interviewees alleged that some tech journalists do not comprehend the Sidewalk technology or are excessively negative about Amazon for the sake of \textquote{clickbait}. Consulted industry press releases generally serve marketing purposes. Next to that, Amazon restricts access to certain documents and development portals to Sidewalk-authorised developers \cite[e.g.][]{amazonSidewalk360n.d.}, and often does not disclose all details of technical implementations. As illustration, it does not become clear precisely what (meta)data Amazon processes, and how, for service optimisation \cite{amazonAmazonSidewalkPrivacy2023, amazontechnologiesAmazonSidewalkSpecification2024b}. The protocol specification also considers a \textquote{Detailed specification of Gateways and the Amazon Sidewalk Cloud}, and interactions between them, outside its scope \cite[][p. 10]{amazontechnologiesAmazonSidewalkSpecification2024b}. 

To mitigate, we triangulated multiple data sources to increase the robustness of the case studies and interviews \cite{rowleyUsingCaseStudies2002, yinCaseStudyResearch2017, solarinoChallengesBestpracticeRecommendations2021}.

\section{Conclusion} \label{sec_conclusion} 
On the surface, applying PETs across millions of devices seems a privacy win, but our empirical case study reveals a more complex story. We started with the hypothesis that tech companies may implement PETs in their computational infrastructure such that they entrench it in the production of other organisations. Our study made concrete how Amazon's implementation of PETs in Sidewalk manifests the two-faceted \emph{PET Paradox}. Leveraging PETs and control over consumer devices, Amazon clips a part of an infrastructure of internet-connected devices (gateways; managed by consumers and ISPs) for a closed network that hooks IoT devices (endpoints of manufacturers) onto their cloud. The PETs conceal some information flows, but also enable Amazon to create new ones, expanding its production environment. For people, the new flows compromise privacy; for manufacturers, their confidentiality and competitive position vis-à-vis Amazon. Seen this way, PETs may serve to exacerbate rather than reduce information asymmetry-based power imbalances (facet 1). 
	
Additionally, Amazon influences the production of manufacturers adopting this cloud connectivity offering; spanning across hardware, OS, software, and the environments producing and managing their devices and services. With this influence, Amazon turns their devices into accessories of Amazon's cloud, while its governance of Sidewalk eats into companies' technical and organisational bandwidth, incurring a path dependency. Thus, Amazon tailors privacy in a fashion that entrenches their computational infrastructure, and thereby infrastructural power, in the production of other companies. This increases Amazon's capacity to determine the mechanisms needed to remotely manage devices from AWS, impacting the viability of implementing PETs in IoT devices (facet 2). Both outcomes are orthogonal to PETs' original aim of limiting corporate and governmental power. \\
	
\noindent We propose the following paths forward. The first facet of the PET Paradox calls for broadening privacy analyses to elicit novel risks that go beyond questions of confidentiality of user data. The second facet is harder to tackle. Scholars and governments are actively analysing the digital advertising and finding network instances as gatekeeping and self-preferencing, and therefore potentially anticompetitive. The digital contact tracing instance raised governmental sovereignty and decision-making questions: Google and Apple became arbiters of privacy and obtained an important role in public healthcare. From a policy perspective, privacy, competition, and sovereignty lenses aim to address power asymmetries. However, we need greater discussion on whether and how well they do so, given that they leave tech companies' incursion into organisations' production mostly out of view. We believe governance and implementation of PETs, and their impacts on production of hardware and services, to be fruitful areas for future research. Overall, assessing privacy claims made by companies or governments using PETs should include a broader (technical, economic, and political; B2B and B2G) power analysis beyond privacy (e.g. data minimisation; B2C). Similarly, we call for deliberation about legally enshrining more user control over devices, such that manufacturers cannot silently co-opt them to deliver services to others. 

From a technical perspective, we call for the PETs community to look more critically at dependencies of users and manufacturers on tech companies' computational infrastructures. As \citet{troncosoSystematizingDecentralizationPrivacy2017} show, many decentralised systems designed to enhance privacy, (sometimes tacitly) assume centralisation of network information and computations, and of trust establishment. These centralisations may be technical (e.g. key exchange, network information) and social (e.g. reputation and identity management). These centralising forces -- and the power dynamics they create between users, manufacturers, and tech companies -- need greater attention from PETs researchers. In Sidewalk, we see Amazon's centralising forces in networking (e.g. managing network topology, gateways, routing, authentication, usage limits, abuse mitigation, and key exchange), its cloud environment (e.g. funnelling manufacturers into AWS, and defining how AWS services work), and in manufacturers' production (e.g. certification, hardware, and software update requirements). Cloud communication already seems prevalent in the IoT, underlining the significance of this control \cite[e.g.][]{schmidtIoTFlowInferringIoT2023}. This also applies to Android messaging apps, that rely on and may leak personal information to Google's and Apple's push notification services \cite{samarinMediumMessageHow2024}.

The remote and increasingly centralised management of privacy and security is a field deserving of more attention. The impact of mechanisms to remotely manage and update devices -- often as part of cloud-based production environments -- on the adoption of PETs is understudied, given their prevalence. Part and parcel to Amazon, Apple and Google \textquote{clipping} computational power and connectivity of consumer devices, is their ability to remotely transform them. With Sidewalk, Amazon even creates a novel connectivity channel for themselves and manufacturers to ship new functionality to and collect telemetry from endpoints, without user intervention. While in theory good for device security and functionality -- reducing reliance on users to patch their devices, and easing maintenance of devices in remote locations -- Amazon's computational infrastructure becomes entrenched in the way that these devices operate. 

Concretely, examples exist that decentralise (some of) the above-mentioned forces. For instance, \citet{ashurPrivacyPreservingDeviceTracking2018} present a decentralised privacy-preserving asset tracking solution. \citet{groschuppItsTEEtimeNew2023} propose a trusted execution-based smartphone architecture that curbs Apple and Google's ability to constrain device functionality through their OSes. This design reduces the need for user trust in OSes regarding privacy, while making \textquote{users sovereign over their phones} (p. 1) and empowering them, phone manufacturers, and app developers. We encourage more such research to scrutinise centralising forces such as OTA infrastructures, OSes, cloud environments, patches and updates, and hardware implementations of public key infrastructures. This research should not merely assess privacy and security, but study how they may avoid entrenching existing computational infrastructures.


\begin{acks}
	This work is part of the Programmable Infrastructures Project\footnote{\url{www.tudelft.nl/en/tpm/our-faculty/departments/multi-actor-systems/research/projects/programmable-infrastructures-project}} and was partially funded by Fondation Botnar and NWO project Public Values in the Algorithmic Society (AlgoSoc, file number 024.005.017)\footnote{\url{www.algosoc.org}}. 
	For fruitful discussions on earlier drafts, we thank
	Agathe Balayn,
	Aniketh Girish,
	Bernd Kasparek,
	Carmela Troncoso,
	Corinne Cath,
	Mark de Reuver,
	Martina Lindorfer,
	Michael Veale,
	Michel van Eeten,
	Narseo Vallina-Rodriguez, 
	Petros Terzis,
	Tobias Fiebig,
	and the participants of the Privacy Law Scholars Conference Europe 2024.    
\end{acks}

\bibliographystyle{ACM-Reference-Format}
\bibliography{refs}

\appendix
\section{Elaboration of methods} \label{appendix_methods}
This appendix further details our methods (§\ref{sec_methods_and_case_description_methods}). We list the identified adopters (§\ref{adopters}), elaborate on the interview and coding process (§\ref{interview_questions}), and present the interview questions (§\ref{appendix-interview-questions}).

\subsection{Overview of identified adopters} \label{adopters}
Table \ref{tab:adopters} displays the Sidewalk adopters identified with the grey literature review. These results were at points supplemented with information obtained during the interviews. To protect participants' anonymity, we make no reference to specific interviewees here. Note that for most companies, catering to business or consumer users is not a fully binary choice. We categorised offerings based on companies' marketing strategies as they appeared in public materials and the interviews. For instance, devices by consumer-oriented brand MerryIoT can also be used in business contexts such as offices, but their marketing signals a focus on consumers.

\begin{table*}[ht]
	\centering
	\footnotesize
	\caption[Overview of identified Sidewalk adopters]{Overview of identified Sidewalk adopters, products, and business orientations}
	\label{tab:adopters}
	\resizebox{\linewidth}{!}{
		\begin{tabularx}{\linewidth}{
				>{\hsize=0.8\hsize}X 
				>{\hsize=2.4\hsize}X
				>{\hsize=0.6\hsize}X 
				>{\hsize=0.6\hsize}X 
				>{\hsize=0.6\hsize}X 
			}
			\toprule
			\textbf{Company name} & \textbf{Product name (if available): functionality} & \textbf{Sells to businesses (B) or consumeres (C)} & \textbf{End-user is business (B) or consumer (C)} & \textbf{References} \\
				\midrule
				Airthings & Sensing CO2, radon, temperature, humidity, and air quality & B2B, B2C & B, C  & \cite{ballanceWaveUnderstandingConnectivity2024, ballanceWhatDifferenceAirthings2024, texasinstrumentsConnectAmazonSidewalk2020a, airthingsAirthingsn.d., airthingsAnnualReport20222023} \\
				Arrive & Point, Bank, Convey, Package Tower: smart mailbox(es), Mailbox as a Service & B2B   & B & \cite{arriveArriveAnnouncesSuccessful2023, arriveArriveCampaignPicMiin.d., arriveArrive2024} \\
				CareBand & CareBand: panic button, location and activity detection; for elderly people, contact tracing, and outdoors worker safety & B2B, B2C & B, C   & \cite{carebandTechnologyn.d., carebandProductsn.d., higginbothamCareBandBetAmazon2023} \\
				DeNova Detect (by New Cosmos) & 807NAS: natural gas alarm & B2B, B2C & B, C  & \cite{newcosmosusaDeNovaDetect807NAS2023, newcosmosusaDeNovaDetectn.d., newcosmosusaAmazonSidewalkSmartn.d.} \\
				Deviceroy & Aria: relaying an industrial device's readings to the internet & B2B   & B     & \cite{deviceroyAriaSpecSheet2023, deviceroyConnectivityRevolutionDeviceroy2023} \\
				Level & Level: smart door lock & B2B, B2C & B, C  & \cite{levelSupportHomeIntegrationsn.d., levelMultifamilyn.d., levelSmartLockCatalogn.d., amazonEchoTileLevel2021} \\
				MerryIoT (by Browan Communications) & Four devices for sensing CO2, motion, door/window open/close, water leak, temperature, and humidity & B2C   & C     & \cite{merryiotProductn.d., merryiotMerryIoTSensorsMerryIoT2023} \\
				Meshify (by HSB) & Defender S: water leak and water pipe freeze/break sensing & B2B   & B      & \cite{meshifyn.d., wightPromisedGrowthIoT2023a, meshifyMeshifyDefendern.d.} \\
				MOKO-Smart & Motion detection, asset and person tracking, smart plug & B2B (OEM) & B, C  & \cite{kuanExtendYourIoT2023} \\
				Netvox & S315 series: integrates modular sensors; supports sensing temperature, humidity, motion, water leaks, vibration, light, and door contact & B2B (OEM) & B, C  & \cite{netvoxNewNetvoxS3152023, netvoxProductsAmazonSidewalkn.d., amazonAmazonInvitesDevelopers2023} \\
				OnAsset & Sentinel 200: asset tracking and condition monitoring)  &  B2B   & B    & \cite{gonsalvesOnAssetLogisticsService2023, onassetintelligenceOnAssetIntelligenceLaunches2023, onassetintelligenceProductsn.d., amazonAmazonInvitesDevelopers2023} \\
				Primax & Woody: smart door lock & B2B (OEM) & B, C   & \cite{amazonAmazonInvitesDevelopers2023, primaxelectronicsPrimaxElectronicsLtdn.d.} \\
				Subeca & Pin: Advanced Meter Infrastructure sensor for water utilities & B2B   & B  & \cite{subecaSubecaIncJoins2023, smartwaterwatchSubecaAimsBring2023, subecaEnhancingWaterUtility2023} \\
				Tag-n-Trac & Smart Sense: asset tracking and condition monitoring & B2B   & B  & \cite{tag-n-tracSmartSensingn.d., texasinstrumentsConnectAmazonSidewalk2020a} \\
				Thingy & Thingy: air quality sensing, specifically for early wildfire detection & B2B   & B  & \cite{wallerThingySidewalkWildfire2022, thingy-iotHomen.d.} \\
				Tile  & Tile: asset tracking & B2B, B2C & B, C  & \cite{amazonEchoTileLevel2021, tileAmazonSidewalkStrengthen2021} \\
				\bottomrule
			\end{tabularx}
		}
	\end{table*}

\subsection{Elite interviews: recruitment and coding} \label{interview_questions}
We sent 300-character LinkedIn connection requests to 94 employees of the sixteen Sidewalk-adopting manufacturers. 23 prospects responded and eight ultimately participated. To sustain a balance between the interviewed organisations, we did not interview multiple people from the same company. Of the participants, four hold a C-level position, two are department heads, and two are high-ranking engineers. 

We initially set out to enrich these findings with perspectives from IoT companies that have considered but not adopted Sidewalk, and approached 25 employees of four non-Sidewalk adopting IoT companies in the smart home domain. After receiving only one positive reply, and because there is no \textit{ex ante} telling whether companies even considered Sidewalk, we halted this endeavour; leaving only \icite{N1} as the interviewed non-adopter. 

For each transcript, two coders participated in coding and discussed the findings with the third. Herein, we took a predominantly inductive approach. Concretely, we devised codes during the analysis and predominantly in \textit{in vivo} fashion to capture respondents' own language and sentiment.
For the first cycle, we iteratively combined initial, structure, and process coding; and evaluation coding. The second cycle involved pattern coding and axial coding. We devised 330 codes across a three-tiered coding scheme. Table \ref{tab:codebook} shows the eleven top-level categories and the corresponding second-tier codes. We do not include the lowest-tier codes for maintaining anonymity of participants as well as readability.

\begin{table*}[htp]
	\centering
	\footnotesize
	\caption{Excerpt of codebook, showing the first-tier (in bold) and second-tier codes}
	\label{tab:codebook}
	\resizebox{\linewidth}{!}{
		\begin{tabularx}{\linewidth}{
				>{\hsize=1.7\hsize}X
				>{\hsize=0.3\hsize}X 
				>{\hsize=1.7\hsize}X 
				>{\hsize=0.3\hsize}X 
			}
			\toprule
			\textbf{Code} & \textbf{Grounded} & \textbf{Code (cont.)} & \textbf{Grounded (cont.)} \\
			\midrule
			\textbf{Advantages of Sidewalk} & \textbf{48} & \textbf{IoT industry dynamics, competition, and partnerships} & \multicolumn{1}{c}{\textbf{138}} \\
			A connectivity service without the manufacturer nor end-user needing to put out gateways & 11    & Amazon competing with telecom providers and tech companies & \multicolumn{1}{c}{5} \\
			Cheap & 10    & Competition by Amazon: countermeasures taken & \multicolumn{1}{c}{8} \\
			Developing the LoRa/IoT/smarthome ecosystems and markets & 13    & Competition by Amazon: perceived impact / level of concern & \multicolumn{1}{c}{24} \\
			Improves / eases user experience & 8     & Competition by Amazon: perceived probability & \multicolumn{1}{c}{22} \\
			Improves adopter's reputation / leverages Amazon's reputation & 4     & Competition by Apple and Google & \multicolumn{1}{c}{4} \\
			Perception of (the significance of) Sidewalk's value for company & 10    & Manufacturer's other partnerships & \multicolumn{1}{c}{6} \\
			\textbf{Alternative connectivity protocols} & \textbf{88} & Manufacturer's perception of Sidewalk longevity/sustainability & \multicolumn{1}{c}{3} \\
			Bluetooth & 3     & Relation of Amazon with silicon providers & \multicolumn{1}{c}{11} \\
			Cellular & 1     & Relation of manufacturer with Amazon not regarding Sidewalk (e.g. Marketplace) & \multicolumn{1}{c}{15} \\
			Comcast MachineQ & 3     & Relation of manufacturer with Amazon regarding Sidewalk & \multicolumn{1}{c}{26} \\
			Complicated to support multiple protocols & 3     & Relation of manufacturer with silicon provider & \multicolumn{1}{c}{15} \\
			Experience with supporting multiple protocols simultaneously & 30    & Risk of commoditisation when adopting another company's network/protocol & \multicolumn{1}{c}{3} \\
			Helium & 2     & Vision of competitive ecosystem: comparison with Amazon: "choices ... based on what we knew, is the same as someone who probably knows more" & \multicolumn{1}{c}{2} \\
			LoRa  & 16    & Vision of future Sidewalk/LoRaWAN ecosystem ("I think at the end of the day, LoRaWAN Alliance would prevail") & \multicolumn{1}{c}{10} \\
			LoRa satellite & 2     & \textbf{Privacy} & \multicolumn{1}{c}{\textbf{55}} \\
			Matter & 12    & (Merits of) public privacy concerns of Sidewalk's privacy, and PETs to address them & \multicolumn{1}{c}{18} \\
			Rationale for supporting multiple protocols simultaneously & 21    & Advantage of opt-out roll-out: coverage without user intervention & \multicolumn{1}{c}{6} \\
			Sigfox & 3     & Company's privacy governance: no change after adopting Sidewalk & \multicolumn{1}{c}{7} \\
			UWB   & 1     & PETs and data visibility & \multicolumn{1}{c}{15} \\
			Wi-SUN & 1     & Privacy as a (more or less important) customer requirement / precondition for delivering IoT services & \multicolumn{1}{c}{7} \\
			Wifi  & 3     & Public reception of other crowdsourced services/infrastructures & \multicolumn{1}{c}{4} \\
			Xfinity Connect (Comcast) & 1     & \textbf{Production before and after adopting Sidewalk} & \multicolumn{1}{c}{\textbf{57}} \\
			\textbf{Amazon's culture and motivation to develop and marketing of Sidewalk} & \textbf{46} & (Changes in) rationales for cloud usage & \multicolumn{1}{c}{15} \\
			"everybody wanted to own the smart home" & 2     & Changes in (producing with) chips / endpoint hardware & \multicolumn{1}{c}{3} \\
			Amazon culture and the position of Sidewalk vis-à-vis other services/products/departments & 8     & Cloud usage (AWS) & \multicolumn{1}{c}{23} \\
			Creating an ecosystem / infrastructure / layer for IoT networking & 9     & Cloud usage (non-AWS) & \multicolumn{1}{c}{15} \\
			Expectation of Sidewalk roll-out outside USA & 4     & Experienced difficulty in adopting Sidewalk & \multicolumn{1}{c}{5} \\
			Increase manufacturers' use of AWS & 2     & Factory line, device provisioning, and implementing encryption & \multicolumn{1}{c}{10} \\
			Marketing towards consumers (gateway/endpoint users/owners) & 7     & \textbf{Security} & \multicolumn{1}{c}{\textbf{46}} \\
			Marketing towards manufacturers & 3     & Company's security measures and principles (in addition to Sidewalk's / before adopting) & \multicolumn{1}{c}{14} \\
			Obtaining "data" & 2     & Competing on security with other IoT companies & \multicolumn{1}{c}{5} \\
			Potentially useful for logistics department & 1     & Customer's perception / assessment of security for interviewee's services and/or Sidewalk & \multicolumn{1}{c}{5} \\
			Reduction of Amazon's hardware business & 9     & Perception of Sidewalk's level of security & \multicolumn{1}{c}{18} \\
			\textbf{Disadvantages of Sidewalk} & \textbf{42} & Specific security requirements for own product/service & \multicolumn{1}{c}{6} \\
			Gateway owners did not consent & 2     & Using cloud services for cybersecurity regulation compliance & \multicolumn{1}{c}{3} \\
			Insufficient coverage & 1     & \textbf{Sidewalk governance} & \multicolumn{1}{c}{\textbf{47}} \\
			Lacking functionality/utility & 4     & Certification process & \multicolumn{1}{c}{9} \\
			Need other protocols as backup/mitigation for reliance & 6     & Closed, proprietary network & \multicolumn{1}{c}{20} \\
			Negative public reception of Sidewalk & 8     & Invite-only early stages of Sidewalk development & \multicolumn{1}{c}{2} \\
			Reliances on and conflicts of interest with Amazon & 11    & Organisational and factory auditing & \multicolumn{1}{c}{7} \\
			Technical and organisational resource constraints (e.g. processing capability, bandwidth) & 11    & Policies/usage requirements & \multicolumn{1}{c}{7} \\
			\textbf{Interviewee company profiles} & \textbf{63} & Watchdog role & \multicolumn{1}{c}{3} \\
			B2B and B2C interplay & 8     & \textbf{Technical architectures of endpoints} & \multicolumn{1}{c}{\textbf{25}} \\
			Endpoint use cases and company's domains of activity & 12    & Endpoint architectures, functionalities, and integrations & \multicolumn{1}{c}{22} \\
			Interviewee's path to adopting Sidewalk & 23    & Utilisation of Sidewalk architecture (e.g. mostly up- or downlink traffic) & \multicolumn{1}{c}{3} \\
			Rationales for B2B orientation & 20    &       &  \\
			Rationales for B2C orientation & 18    &       &  \\
			\bottomrule
		\end{tabularx}
	}
\end{table*}

\subsection{Interview questions} \label{appendix-interview-questions}
Below is a compilation of the prepared interview questions. Because of the semi-structured nature, and because some questions were not applicable for all respondents, we may have skipped certain questions or formulated them differently. Likewise, and in order to ensure spontaneous answers, we did not share interview questions beforehand -- except in one case where the participant requested us to. We only provided the interviewees with the clarifications appended to the question between parentheses if they did not understand the question, to prevent inducing bias in their answer. Further personal questions used for rapport building are not included to protect anonymity of the participants.

\subsubsection{Rapport building} \hfill \break
\noindent \emph{Background and expertise of interviewee}
\setlist{nolistsep} 
\begin{enumerate}
	\item Can you tell me about your company and your role therein? \hfill \break
\end{enumerate}

\noindent \emph{Company profile}
\begin{enumerate}[resume]
	\item What kind of customers do you cater to, and which has your primary focus? \emph{(E.g. business or consumer users; geographical area of focus; market domain (building management, logistics, utilities))}
	\item What share of your customers is business or consumer?
	\item What drew you to catering to these types of customers?
	\item What are the differences in the requirements from and use cases of the different customer types that you cater to?
	\item What type of customer do you think Sidewalk has a stronger case for? \emph{(E.g. B2C or B2B)}
\end{enumerate}

\subsubsection{Grand tour and mini-tour questions} \hfill \break
\noindent \emph{Adoption and motivation}
\begin{enumerate}[resume]
	\item How did you discover Sidewalk?
	\item What novel opportunities does Sidewalk provide? \emph{(E.g. compared to other connectivity protocols)}
	\item How important is Sidewalk to your product, service, or company? \emph{(E.g. compared to the other connectivity modes you use)}
	\item Is your service really based on the availability of Sidewalk, or were you already developing your service before Sidewalk was published?
	\item Which of the three connectivity protocols of Sidewalk do you use (i.e. LoRa, FSK, BLE)?
	\item What connectivity methods other than Sidewalk have you considered? \emph{(E.g. LoRaWAN, Bluetooth, Matter)}
	\item How does Sidewalk compare to these other methods?
	\item I found references mentioning your company’s (prospective) adoption of Sidewalk, but your product page does not mention Sidewalk and lacks the `Works with Amazon Sidewalk’-badge. What is your adoption status?
	\item Your organisation sells both Sidewalk-compatible devices, and non-compatible counterparts using other connectivity methods. Why the separation?
	\item Did you have doubts around adopting Sidewalk? How were they addressed, or what pulled you over the line? \\
\end{enumerate}

\noindent \emph{Privacy and security}
\begin{enumerate}[resume]
	\item Has your use of Sidewalk changed the privacy architecture or governance of your IoT offerings? 
	\item Does Sidewalk help address privacy and/or cybersecurity concerns better than other communication methods? \\
\end{enumerate}

\noindent \emph{Production and cloud usage}
\begin{enumerate}[resume]
	\item Does your company use the cloud for offering or producing your products and services? If so, which provider do you use?
	\item Were you already using cloud services before adopting Sidewalk?
	\item Has adopting Sidewalk led to a change in how you use the cloud? If yes, how?
	\item Did your earlier use of AWS ease your adoption of Sidewalk?
	\item How do you process data sent over Sidewalk?
	\item How do you use AWS IoT Core for Sidewalk?
	\item How do you process data sent over other connectivity protocols? \emph{(E.g. LoRaWAN)}
	\item Is the data sent over Sidewalk processed differently than data sent over other connectivity protocols, such as LoRaWAN?
	\item Has adopting Sidewalk changed the way in which you produce your devices? If yes, how? \emph{(E.g. with relation to key management or enabling device authentication with Sidewalk; or by enabling remote updating of endpoints or having them send more telemetry)}
	\item Who is your silicon provider? What is your collaboration with them like, also with regards to adopting Sidewalk? \\
\end{enumerate}

\noindent \emph{Governance}
\begin{enumerate}[resume]
	\item Have you encountered any policies and/or requirements that your product or organisation is subject to because of using Sidewalk?
	\item I saw that there is a quite elaborate process on how to get your devices Sidewalk-certified. How did this process go for you?
	\item Can you elaborate on the relation between the LoRa Alliance and Amazon? \\
\end{enumerate}

\noindent \emph{Information flows to and competition with Amazon}
\begin{enumerate}[resume]
	\item I imagine there might be usage data that gives away how your device functions. For example, how big are the payloads; how quick does the battery deplete; how often does the device communicate with the cloud. What information do you think Amazon is able to see about your devices and cloud use?
	\item Do you think Amazon could use this information to improve their own offerings? \emph{(E.g. their own IoT devices, or their AWS services)}
	\item Do you have insights into whether Amazon is developing new endpoints themselves? \\
\end{enumerate}

\noindent \emph{Reliance}
\begin{enumerate}[resume]
	\item How would your organisation or service be affected if Amazon were to pull the plug on Amazon, or if your partnership falls through?
\end{enumerate}

\subsubsection{Closing questions}
\begin{enumerate}[resume]
	\item Is there anything that you expected me to ask that I have not, or anything else that you would like to share?
	
	\item Now that you understand what kind of questions and subjects interest me, and given your expertise in the field, are there other people that you think I should talk to, in your organisation or broader network?
\end{enumerate}

\section{Sidewalk marketing visuals} \label{appendix_marketing_visuals}
\begin{figure}[ht]
	\centering
	\includegraphics[width=\linewidth]{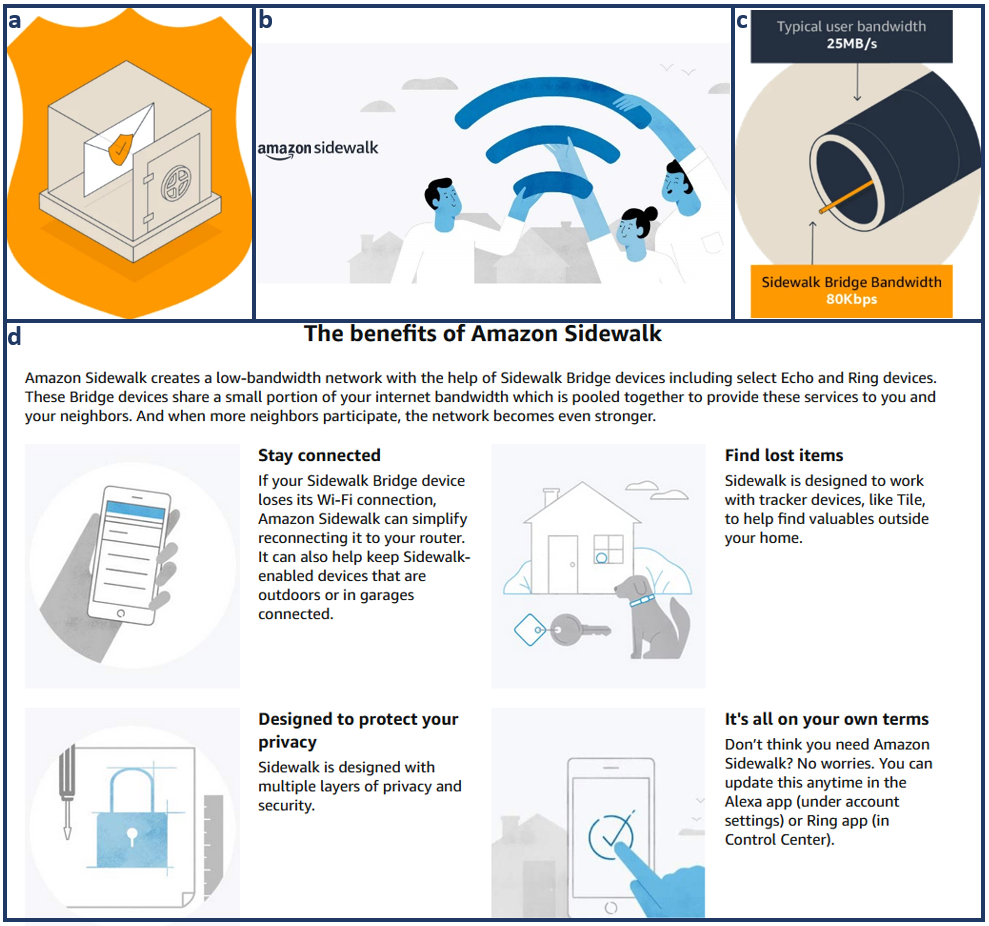}
	\caption[A compilation of visuals that Amazon uses to stress Sidewalk's multi-layered encryption, crowdsourced nature benefiting communities, bandwidth constraints, and use cases]{A compilation of visuals that Amazon uses to stress Sidewalk's multi-layered encryption \textit{(a, d)}, crowdsourced nature benefiting communities \textit{(b, d)}, bandwidth constraints \textit{(c)}, and use cases \textit{(d)}. 
		\textit{a} and \textit{c} reproduced from \cite{oneillEverythingYouNeed2023a};
		\textit{b} and \textit{d} reproduced from \cite{amazonAmazonSidewalkn.d.}.
		}
	\Description{A compilation of visuals that Amazon uses to stress Sidewalk's multi-layered encryption, crowdsourced nature benefiting communities, bandwidth constraints, and use cases}
	\label{fig:marketing}
\end{figure}

Figure \ref{fig:marketing} shows examples of how Amazon visualises Sidewalk to consumers (both gateway and endpoint owners). 

\section{Elaboration of information flows and PETs in Sidewalk} \label{appendix_information_flows_and_PETs}
This appendix supplements the analysis of information flows and PETs in Sidewalk (§\ref{sec_information_flows_and_PETs_in_sidewalk}). We first discuss privacy concerns of Echo and Ring devices irrespective of Sidewalk (§\ref{echo_ring_privacy_concerns_regardless_of_sidewalk}). Next, we elaborate on Sidewalk's PETs (§\ref{sidewalk_PETs}). This allows us to outline which privacy concerns they leave unaddressed (§\ref{new_or_exacarbated_privacy_concerns}), and which telemetry information flows they do not prevent (§\ref{endpoint_telemetry_visibility}).

\subsection{Echo and Ring privacy concerns regardless of Sidewalk} \label{echo_ring_privacy_concerns_regardless_of_sidewalk}
We distinguish three types of surveillance that Echo and Ring devices enable, that infringe individual privacy. First is \textit{peer surveillance}, here defined as consumers using Echo and Ring devices to surveil their peers, e.g. neighbours, partners, or family members. Amazon's Ring brand offers an extensive suite of surveillance products and services, including video doorbells, cameras, and security systems \cite{ringProductsn.d.}. Ring and Echo users can monitor their environments, including family, neighbours, passers-by, and workers, that are generally unable to avoid this monitoring \cite{kellyRingVideoDoorbell2022, selingerAmazonsRingSurveillance2022,		stephensonAbuseVectorsFramework2023}. In fact, Amazon has rewarded citizens with discounts or free products when grouping up in Ring \textquote{Digital Neighborhood Watches} and reporting crime \cite{haskinsRingToldPeople2019c}. \citet{nguyenDigitalDoorstepHow2022a} argue that Ring lets Amazon convert a labour cost of monitoring their delivery drivers and parcels themselves, into a source of income: Ring users can monitor deliverers with cameras and sanction them through reviews or sharing recordings on social media. Other scholarship outlines how Ring surveillance grows racial injustices, and that Ring users tend to over-exaggerate how suspicious or criminal a passer-by is \cite{calacciCopYourNeighbor2022}. 
 
Second is \textit{customer surveillance}, i.e. Amazon monitoring customers' devices, as well as how people interact with them. Numerous articles outline how Amazon logs interactions with their devices to improve their own services or offer personalised services and advertisements \cite[e.g.][]{kellyRingVideoDoorbell2022, westAmazonSurveillanceService2019, iqbalTrackingProfilingAd2023}. A report from a cybersecurity company found Amazon's IoT device companion apps to be the most \textquote{data-hungry} of the 290 apps they studied \cite{surfsharkMethodology2024, surfsharkInsightsOurSmart2024}. Further, Amazon has a history of giving employees too liberate access to Ring users' videos and Echo voice recordings, without informing users thereof \cite{brodkinFTCAmazonRing2023, pattersonHowKeepAmazon2019}. In this context, we also point out Ring's history of cybersecurity issues \cite[e.g.][]{ngRingDoorbellsHad2019c, guarigliaWhatKnowYou2020a, guarigliaAmazonRingEndtoEnd2021}. 

Third, Echo and Ring devices stimulate \textit{law enforcement surveillance}, that Amazon actively pursues through extensive partnerships with police departments \cite[e.g.][]{soWhyWeDont2023}. Echo devices could aid police investigations \cite{orrAlexaDidYou2018} and US law enforcement set out to insert voice recordings of an Echo in a criminal court case in 2017 \cite{jacksonStudySecurityPrivacy2018}. On top of that, Amazon has clear ambitions to offer access to its country-wide surveillance network to law enforcement. Reportedly, Amazon counts over 2,000 US police and fire departments as their partners \cite{lyonsAmazonsRingNow2021}; for instance to have law enforcement hand out devices to citizens for free and to train officers in PR and handling press questions \cite{haskinsAmazonRequiresPolice2019c}. Amazon also persuaded municipalities to subsidise residents' purchases of Ring products \cite{haskinsUSCitiesAre2019c}. Resultingly, Ring cameras are widespread throughout the US. Amazon provided authorities a map of Ring devices and, until February 2024, an easy way to request footage from a camera's owner that Ring owners could approve in a smartphone app, without requiring a warrant \cite{lyonsAmazonsRingNow2021}. Even though Amazon has now removed this button, authorities can still obtain footage with a warrant or by demonstrating to Amazon that they need it for an ongoing emergency \cite{dayAmazonRingStop2024}. Indeed, Amazon has disclosed video feeds without warrants to authorities based on their own \textquote{good-faith determination} \cite[][p. 4]{amazonAmazonResponseSenator2022a}, a practice that some authors rightfully take issue with \cite[e.g.][]{guarigliaVictoryRingAnnounces2024}. The sensitivity of these requests is illustrated by the Los Angeles police requesting Ring feeds capturing Black Lives Matters protests \cite{guarigliaLAPDRequestedRing2021a}. Meanwhile, numerous reports refute Ring's alleged contribution to combating crime \cite{harrisVideoDoorbellFirm2018a, farivarCuteVideosLittle2020a, guarigliaWhatKnowYou2020a}.

\subsection{Elaboration of Sidewalk's PETs} \label{sidewalk_PETs}
This section briefly outlines how Sidewalk's two PETs work [\citealp[][]{amazonAmazonSidewalkPrivacy2023}; \citealp[][]{amazonComponentsAmazonSidewalk2023}; \citealp[][chapter 4]{amazontechnologiesAmazonSidewalkSpecification2024b}]. Sidewalk features a three-layer end-to-end \emph{encryption scheme}. This scheme encrypts both payload data and certain metadata from being visible to and tampered with by others. To illustrate: this works as follows for an endpoint sending information to an application server. The \textit{endpoint} encrypts the payload data with an Application Server Key (only known to the endpoint and application server), and encrypts the result and the network layer frame with a Sidewalk Network Server Key (only known to the endpoint and Sidewalk Network Server). The endpoint then sends the packet to the \textit{gateway}, that adds a third layer using the Gateway Network Server Key (only known to the gateway and Sidewalk Network Server). The \textit{network server} then decrypts the second and third layer and forwards the packet to the appropriate \textit{application server}, that decrypts the final layer and processes the payload. Accordingly, manufacturers must embed appropriate certificates and key pairs into endpoints during their production. Downlink traffic is secured similarly.

Second is \emph{device identifier obfuscation}. For the encryption and authentication, all endpoints and gateways must carry unique credentials. To \textquote{minimize data tied to customers} \cite[][pp. 11-12]{amazonAmazonSidewalkPrivacy2023}, Amazon uses temporary identifiers derived from the unique credentials for some functionalities. For instance, transmission and gateway identifiers (which are different than the device's persistent identifier, usually its serial number) are renewed every fifteen minutes and information used for routing packets cleared every 24 hours.

\subsection{Privacy concerns that Sidewalk raises or exacerbates} \label{new_or_exacarbated_privacy_concerns}
Despite its PETs, Sidewalk compromises user privacy by facilitating surveillance (§\ref{new_or_exacarbated_privacy_concerns_surveillance}), not providing insight into the security mechanisms (§\ref{new_or_exacarbated_privacy_concerns_security_transparency}), blurring privacy boundaries (§\ref{new_or_exacarbated_privacy_concerns_boundaries}), and undermining owner control over their devices (§\ref{new_or_exacarbated_privacy_concerns_opt-out}).

\subsubsection{Surveillance} \label{new_or_exacarbated_privacy_concerns_surveillance}
While Echo and Ring devices already enabled three types of surveillance, Sidewalk expands the area wherein Echo and Ring devices can work, because they are both a gateway and endpoint simultaneously. They can thus be placed outside the range of the owner's wifi router, that previously constituted a technical boundary circumscribing where these devices could be used. Sidewalk functions as a thread connecting smaller patches of surveillance products into one great surveillance network across entire neighbourhoods and cities \cite{hanleyEyesEverywhereAmazon2021c}. Consequently, Sidewalk amplifies these already existing privacy issues; but also adds to them.

Regarding \textit{peer surveillance}, among Sidewalk-enabled devices are asset trackers that stalkers can misappropriate \cite{callasUnderstandingAmazonSidewalk2021a, vaasAmazonSidewalkPoised2021a}. In fact, two stalking victims filed a lawsuit against Tile and Amazon over Tile's integration with Sidewalk, arguing that Sidewalk's coverage was vital to the stalking by an ex-partner \cite{karabusStalkingVictimsSue2023, smalleyPlaintiffsTileSuit2024}. Asset trackers more generally have been used for stalking and domestic abuse, that technical solutions proposed by their manufacturers fail to robustly tackle \cite{turkCanKeepThem2023, heinrichPleaseUnstalkMe2024}. In theory, \textit{law enforcement} could also use Sidewalk trackers to track suspects; or, as \citet{paceEveryStepYou2023} detail, recover a tracker's previous locations.

Relatedly, the information flows from endpoints to Amazon and manufacturers, and vice versa (Table \ref{tab:endpoint_data_metadata_telemetry}) enable \textit{customer surveillance}. For instance, Amazon can determine endpoints' and thus their users' locations \cite[e.g.][]{cristAmazonSidewalkWill2021a}, as Amazon can see which gateways endpoints connect to \cite{despresWhereSidewalkEnds2022}. According to \citet{despresWhereSidewalkEnds2022}, the Whitepaper \textquote{claims to forget the device ID associated with a transmission after replacing it with a temporary rotating identifier. In reality, ... device IDs are kept to enable bidirectional communication, as the most likely gateway to still be in communication with the device is the one that handled its last transmission} (p. 3). To be more precise, Sidewalk identifies endpoints with one of three identifiers, namely the Sidewalk Manufacturing Serial Number (before an endpoint’s registration to Sidewalk), Sidewalk ID (loaded into the device during registration), and the Transmission Identifier (after registration) \cite[][pp. 49-50]{amazontechnologiesAmazonSidewalkSpecification2024b}. Amazon also uses these persistent identifiers to block endpoints from accessing the network if these are reported as lost, suffer from security issues, or \textquote{if a third party [manufacturer] fails to act in good faith} \cite[][p. 14]{amazonAmazonSidewalkPrivacy2023}. Manufacturers can similarly track their users, as Amazon acknowledges in its Whitepaper: \textquote{Third-party Sidewalk device manufacturers may maintain their own logs that are subject to their respective retention periods and privacy notices} \cite[][p. 6]{amazonAmazonSidewalkPrivacy2023}.

\subsubsection{Security and lack of transparency} \label{new_or_exacarbated_privacy_concerns_security_transparency}
Some authors remark that novel technologies are rarely bug-free; Sidewalk's quick opt-out roll-out and resulting large coverage could mean that security flaws are only uncovered when it is already widely used \cite{callasUnderstandingAmazonSidewalk2021a, vaasAmazonSidewalkPoised2021a}. Amazon has not published details about how precisely it implements its security measures in Sidewalk, nor enabled independent reviews \cite{callasUnderstandingAmazonSidewalk2021a}. This pits Amazon as a single point of failure and increases the need for users to trust them, contrary to the rationale of PETs.

\subsubsection{Blurring privacy boundaries} \label{new_or_exacarbated_privacy_concerns_boundaries}
Sidewalk tightens Amazon’s grip over citizens’ personal households and physical livelihoods. Sidewalk differs from conventional mental images of the internet, granted that Sidewalk endpoints are not connected by virtue of their owner's router, but by other people's gateways (and their routers) and the Sidewalk Network Server \cite{chattingDesignReappearanceSmart2021a}. With connectivity enabling digital services that reach beyond the range of one’s home router(s), i.e. into the yard or streets, Sidewalk gateways further blur the borders between private and public spaces around people’s homes \cite{humphryVisibilitySecuritySmart2021a}, scaling smart homes up to smart neighbourhoods \cite{cristAmazonSidewalkWill2021a} that Amazon mediates. Thus, Sidewalk is not a digital service confined to cyberspace, but invades public and private physical spaces \cite{cleaveTwoSunsData2021a}. 

\subsubsection{Undermining owner control over devices} \label{new_or_exacarbated_privacy_concerns_opt-out}
A prominent objection to Sidewalk by journalists and advocates is to Amazon's opt-out and relatively silent transformation of Echo and Ring devices into Sidewalk gateways. This undermines owners' control over their devices, and puts them at risk of violating the terms of service agreements with their ISPs \cite[e.g.][]{chaseAmazonSidewalkWill2021b, callasUnderstandingAmazonSidewalk2021a, pattersonWelcomeAmazonSidewalk2021, newmanHowTurnAmazon2021, goodinAmazonDevicesWill2021}. We question how many users will have seen the notification in their Echo and Ring apps that announced Sidewalk, and how often they open these apps in the first place. Similarly, not all Echo owners might be inclined to open an email titled \textquote{Echo Update: Amazon Sidewalk is coming soon} \cite{buddAmazonSidewalkRollout2020}, which does not mention that a crowdsourced networking service is approaching in opt-out nature. In line with research demonstrating that people tend not to deviate from default settings and that tech companies actively hide privacy settings \cite[e.g.][]{acquistiPrivacyHumanBehavior2015, acquistiEconomicsPrivacy2016a}, authors speculate that Amazon deployed Sidewalk in opt-out fashion to increase gateway participation and thus coverage \cite{callasUnderstandingAmazonSidewalk2021a, goodinAmazonDevicesWill2021, newmanHowTurnAmazon2021, pattersonWelcomeAmazonSidewalk2021}. 

On top of that, the Sidewalk rollout changes the relationship between gateway owners and their device to launch the Sidewalk service. Firstly, Sidewalk gateways will by default share connectivity with endpoints owned by other consumers: something that the gateway owner likely did not envision when bringing the device into their home. Secondly, third-party IoT manufacturers adopting Sidewalk predominantly target business customers. Accordingly, the Echo and Ring devices that consumers have bought, may be used for business or -- in the case of the utilities domain -- public purposes. By stitching together small bandwidth contributions from gateway owners across the US, Amazon constitutes an additional connectivity infrastructure that it allocates to manufacturers. Manufacturers may, in turn, deliver wholly novel services to their customers that gateway owners may not foresee.

\subsection{Endpoint data and telemetry visibility for Amazon and manufacturers} \label{endpoint_telemetry_visibility}
Table \ref{tab:endpoint_data_metadata_telemetry} lists what endpoint data and telemetry is visible for Amazon and to manufacturers. None of this data is visible to gateway owners. Where applicable, Amazon specifies the metadata per connectivity type (e.g. specifying the number of connection attempts individually for BLE, FSK, and LoRa). Metadata about radio frequency broadcasting or signal strength pertains to endpoint-gateway connections. The table is based on Amazon documentation 
[\citealp[][]{amazonAmazonSidewalkPrivacy2023}; \citealp[][]{amazonComponentsAmazonSidewalk2023}; \citealp[][pp. 33-40, 42, 49-55, 60, 175]{amazontechnologiesAmazonSidewalkSpecification2024b}; \citealp[][]{amazonwebservicesAddYourDevicen.d.}] and interviews. Where the references were not specific or complete, we made assumptions (\textquote{Presumably yes}) or provide no judgment (\textquote{Unknown}).

\begin{table*}[htbp]
	\centering
	\footnotesize 
	\caption{Visibility of endpoint data and telemetry for Amazon and manufacturers.}
	\label{tab:endpoint_data_metadata_telemetry}
	\resizebox{\linewidth}{!}{
			\begin{tabularx}{\linewidth}{
					>{\hsize=2\hsize}X 
					>{\hsize=0.4\hsize}X
					>{\hsize=0.6\hsize}X 		
				}		
				\toprule
				\textbf{Endpoint data} & \textbf{Visible to Amazon} & \textbf{Visible to manufacturer} \\
				\midrule
				\multicolumn{3}{l}{\textit{\textbf{Persistent device characteristics}}} \\
				Manufacturer & Yes   & Yes \\
				Serial number (generated by manufacturer) & Yes   & Yes \\
				Advertised Product ID (APID; generated by AWS for manufacturer) & Yes   & Yes \\
				Sidewalk Manufacturing Serial Number (hash of serial number, APID, and device serial number or (if absent) a universally unique identifier) & Yes   & Yes \\
				Sidewalk ID (might be the device’s serial number) & Yes   & Yes \\
				Device profile (DeviceTypeId; created by manufacturer in AWS IoT Wireless to specify the device capabilities) & Yes   & Yes \\
				Qualification process identifier & Yes   & Yes \\
				Associated application server identifier & Yes   & Yes \\
				Supported connectivity protocols (BLE, FSK, LoRa) and parameters (e.g. preferred protocol, (a)synchronous) & Yes   & Yes \\
				Power source (battery- or line-powered) & Yes   & Yes \\
				Maximum transmit power per protocol & Yes   & Yes \\
				\multicolumn{3}{l}{\textit{\textbf{Variable metadata}}} \\
				Payload (e.g endpoint sensor data) & No & Yes \\ 
				Transmission identifier (rotating hash of Sidewalk ID) & Yes   & Yes \\
				Unspecified \textquote{auxiliary data (device and user-related)} that the \textquote{AWS IoT asset services} provide to the application server \cite[][p. 42]{amazontechnologiesAmazonSidewalkSpecification2024b} & Presumably yes & Yes \\
				Identifier of connected gateway & Yes   & No \\
				Location of connected gateway & Yes   & No \\
				Location of endpoint & Yes (through gateway location) & Yes (if manufacturer adds location tracking capability \cite[e.g.][]{sidhuBuildingTrackTrace2024}) \\
				Battery level & Yes   & Yes \\
				Whether the endpoint is throttling down its traffic rate (can be adjusted by Amazon remotely) & Yes   & Yes \\
				Sidewalk SDK version and supported features & Yes   & Yes \\
				Signal strength  & Yes   & Yes   \\
				\multicolumn{3}{l}{\textit{\textbf{Variable telemetry that Amazon can make the endpoint (not) report to them}}} \\
				Periodicity of sending the below metrics to Amazon (by default once a day) & Yes (and can change this periodicity) & Unknown \\
				Number and frequency of uplink and downlink messages & Yes   & Yes \\
				Average size of uplink and downlink messages & Yes   & Presumably yes \\
				Number of (un)successful connections & Yes   & Unknown \\
				Minimum and maximum time taken for a successful connection & Yes   & Unknown \\
				Number and type of communication failures & Yes   & Unknown \\
				Round trip time for messages & Yes   & Yes \\
				Number of duplicated and retried messages & Yes   & Yes \\
				Number of decryption and authentication errors & Yes   & Unknown \\
				Minimum and maximum time for a successful connection  & Yes   & Unknown \\
				Timeout configured for connecting using a certain protocol & Yes   & Presumably yes \\
				Number of (un)successful registrations & Yes   & Unknown \\
				Number of (un)successful key refreshes & Yes   & Unknown  \\
				Number of frustration-free networking retry attempts & Yes   & Unknown \\
				Number of time sync requests performed and received & Yes   & Unknown \\
				Periodicity of time sync requests & Yes   & Unknown \\
				Difference between time as perceived by endpoint and Sidewalk Network Server & Yes   & Unknown \\
				Number of (un)successful Sidewalk Bulk Data Transfer file transfers  & Yes   & Yes \\
				Whether the endpoint has sufficient memory, storage, and battery to install a Sidewalk Bulk Data Transfer-issued update & Yes   & Yes  \\
				Which of these Amazon-sanctioned metrics the endpoint is currently reporting to Amazon & Yes (and can change which) & Unknown \\
				\bottomrule
			\end{tabularx}
		}
	\end{table*}

\end{document}